\documentclass[numberedappendix,floatfix, twocolumn]{aastex631}

\usepackage{bm}
\usepackage{xspace}
\usepackage{enumitem}
\usepackage{graphicx}
\usepackage{color}
\usepackage{comment}
\usepackage{amsmath}
\usepackage[T1]{fontenc}
\usepackage{natbib}
\usepackage[capitalize]{cleveref}
\usepackage{colortbl}
\usepackage[toc]{appendix}
\usepackage{wrapfig}
\usepackage[normalem]{ulem}

\def\red#1{\textcolor{red}{\textbf{}}}
\def\blue#1{#1}

\shorttitle{Late-Time TDE Disk Evolution}
\shortauthors{Alush et al.}

\begin{document}

\title{How Flat is a Plateau?  Evolution of Late-Time TDE Disks}
\correspondingauthor{Yael Alush}
\author[0000-0003-3471-9817]{Yael Alush}
\email{yael.alush@mail.huji.ac.il}
\affiliation{Racah Institute of Physics, The Hebrew University, Jerusalem, 91904, Israel}
\affiliation{Department of Astronomy, University of Wisconsin, Madison, WI, 53706}
\author[0000-0002-4337-9458]{Nicholas C.~Stone}
\affiliation{Racah Institute of Physics, The Hebrew University, Jerusalem, 91904, Israel}
\affiliation{Department of Astronomy, University of Wisconsin, Madison, WI, 53706}
\author[0000-0002-3859-8074]{Sjoert van Velzen}
\affiliation{Leiden Observatory, Leiden University, Postbus 9513, NL-2300 RA Leiden, the Netherlands}

\begin{abstract}
Late-time light curve plateaus in tidal disruption events (TDEs) are often approximated as flat and time-independent. This simplification is motivated by theoretical modeling of spreading late time TDE disks, which predicts slow light curve evolution. However, if time evolution can be detected, late-time light curves yield more information than previously accessible. In this work, we re-examine late-time TDE data to test how well the flat plateau assumption holds. We use Markov Chain Monte Carlo to estimate the maximum likelihood for a family of theory-agnostic models and apply the Akaike information criterion to find that roughly one third of our sample favors evolving plateaus, one third favors truly flat plateaus, and one third shows no statistically significant evidence for any plateau. Next, we refit the TDEs that exhibit statistically significant plateaus using a magnetically elevated $\alpha$-disk model, motivated by the lack of clear thermal instability in late time TDE light curves. From these model-dependent fits, we obtain estimates for the supermassive black hole (SMBH) mass, the mass of the disrupted star, and $\alpha$. Fitted $\alpha$ values \red{range from $10^{-3}$ to $0.4$ (} \blue{have a} mean \red{$\alpha=10^{-1.8}$} \blue{$\alpha=10^{-1.5}$}, with scatter of \red{$0.6$} \blue{1} dex\red{)}, broadly consistent with results from magnetohydrodynamic simulations\blue{, albeit with some outliers}. Finally, we estimate the timescales of disk precession in magnetically elevated TDE models.  Theoretically, we find that disk precession times may be orders of magnitude shorter than in unmagnetized Shakura-Sunyaev disks, and grow in time as $T_{\rm prec}\propto t^{35/36}$; empirically, by using fitted $\alpha$ parameters, we estimate that late time disks may experience $\sim$few-10 precession cycles.

\end{abstract}


\section{Introduction}
\label{sec:intro}

The theory of astrophysical accretion disks developed in the 1970s \citep{PringleRees1972, ShakuraSunyaev1973, LightmanEardley1974, ShakuraSunyaev1976} in response to new observations of accreting black holes \citep{PrendergastBurbdge1968, LyndenBell1969}.  Most conventions and many ideas of analytic and semi-analytic accretion theory date to that time period, from the Shakura-Sunyaev $\alpha$-ansatz to the hysteresis diagrams used to study time-dependent behavior.  While \citet{ShakuraSunyaev1973} disk theory and its descendants have found broad applications in astrophysics, many questions first asked 50 years ago remain unsettled.  How do disks transport their angular momentum?  Are disks viscously and thermally stable when radiation pressure exceeds gas pressure?  What happens when the accretion rate through an accretion disk exceeds the Eddington limit?

These questions remain open for a confluence of reasons.  While modern numerical simulations have demonstrated \citep{BalbusHawley1998} that angular momentum transport is typically governed by the magneto-rotational instability \citep{Velikhov1959, BalbusHawley1991}, this finding has strongly constrained further progress: self-consistent time evolution of theoretical models requires 3-dimensional magnetohydrodynamic (MHD) simulations, which are so costly that \red{they can only achieve} \blue{until recently they have only achieved} inflow equilibrium over a narrow radial range \citep{DeVilliers+2003, Penna+2010, Jiang+2019, Zhang+2025}\blue{, although the development of novel approximations such as cyclic refinement-derefinement loops \citep{Guo+25} may greatly extend such ranges}.  Observations of real accretion disks can offer insight beyond the time horizons covered by \blue{end-to-end} simulations, but are ultimately limited in their ability to answer the aforementioned questions.  The active galactic nuclei that surround supermassive black holes (SMBHs) are well-studied, but generally have sub-Eddington accretion rates \citep{Aird+2018} and (at most radii) viscous times far in excess of a human lifetime.  In X-ray binaries (XRBs), viscous evolution can occur much more rapidly, and indeed these systems provide evidence that some uncertain piece of physics usually suppresses the thermo-viscous instabilities of $\alpha$-disk models \citep{Done+2007}.  However, XRBs are sufficiently complicated that a single $\alpha$ value seems inadequate to capture their angular momentum transport efficiencies \citep{Dubus+2001}, and they often arrive at accretion states very unlike those of simpler disk models \citep{Fender+2003}.

In recent years, the accretion disks of tidal disruption events (TDEs) have been observed in large numbers.  TDEs ensue when stars in galactic nuclei are torn apart by SMBHs \citep{Hills1975, Rees1988, Rossi+2021}.  After a complex and still poorly understood process of circularization \citep{Hayasaki+2013, Guillochon+2014, Shiokawa+2015, Hayasaki+2016, BonnerotLu2020, BonnerotStone2021, Steinberg&Stone2024}, the initially eccentric stellar debris should settle into a relatively simple, axisymmetric, viscously spreading disk.

Observations show that after a fast rise followed by a gradual decay \citep{Gezari+2012, vanVelzen+2021}, the optical and ultraviolet (UV) light curves of TDEs flatten \citep{vanVelzen+2019b}, indicating the transition from complex early-time hydrodynamics to a potentially simple disk-dominated phase. The relative simplicity of late-time TDE emission provides an opportunity to address some of the key open questions in accretion theory outlined above. Furthermore, fitting theoretical models to late-time observations enables us to measure fundamental SMBH parameters, such as SMBH mass and spin \citep{Wen+2020, Mummery&Balbus2020, Wen+2021, Wen+2023, Cao+2023, Mummery+2024, Mummery+2025}. 

However, current theoretical models used for late time TDE disk evolution are {\it ad hoc} in various ways. In principle, time evolution can be determined from the standard Shakura-Sunyaev model, but this relies on the $\alpha$ parametrization of stress: in reality, $\alpha$ is quite uncertain, but unlikely to be a constant \citep{Penna+2013}. Moreover, in regimes where radiation pressure dominates over gas pressure, $\alpha$-disk theory predicts thermo-viscous instabilities \citep{Shen&Matzner2014, Piro&Mockler2025} that are not evidenced in most \citep{vanVelzen+2019b} late-time TDE observations. Attempts to stabilize theoretical models of TDE disks typically invoke modified viscosity prescriptions, which are often not physically motivated. In some cases, late-time modeling is reduced to assuming a pure plateau with no time evolution.

In this paper, we aim to test this assumption by examining how many late-time observations show a flat plateau compared to those that show time evolution. We further fit a late-time disk model that features time evolution and remains both thermally and viscously stable due to a physically motivated picture of magnetic pressure dominance \citep{Begelman&Pringle2007, Kaur+2023, Alush&Stone2025}. From this model, we obtain an estimate of the time-averaged stress to pressure ratio (the Shakura–Sunyaev $\alpha$ parameter). 
The structure of the paper is as follows. In \S \ref{sec:data}, we present the late-time observational sample. In \S \ref{sec:pheno}, we fit phenomenological models to quantify the fraction of TDEs that display a flat versus an evolving plateau and in \S \ref{sec:models} we apply a more physically motivated disk model to extract estimates of the SMBH and disk parameters, including $\alpha$. In \S \ref{sec:astro}, we discuss the astrophysical implications of our results, and in \S \ref{sec:conclusions}, we summarize our findings.

\section{Late-Time TDE Observations}
\label{sec:data}

We use optical and UV observations collected from the \verb|manyTDE| repository\footnote{https://github.com/sjoertvv/manyTDE}, in particular from version 0.6 of this catalog.  For full details on the data reduction, we refer to \citet{vanVelzen+2019,vanVelzen+2021,Mummery+2024}. For each TDE, we correct the difference photometry for Milky Way extinction. \blue{Our fiducial models do not account for dust extinction; however, we perform a sanity check using the extinction measurements from \citet{Guolo+2025} and assess their effect on the inferred plateau temperature. We find a mean change in temperature of $10^{0.04}$ K, with a standard deviation of $10^{0.06}$ K. The largest change is found for AT2021ehb, for which the temperature changes by $10^{0.2}$ K.} As our approach to disk modeling in this paper will be Newtonian, we do not consider X-ray observations of TDEs in this work\footnote{X-ray emission will also be more sensitive to the effects of weak Comptonization \citep{Shimura&Takahara1995, Wen+2020}.}, which require a more careful treatment of relativistic effects \citep{Wen+2020, Mummery&Balbus2020}.

Our initial sample of all TDEs from the repository numbers 98.  We downselect to 45 based on the availability of late time photometry and the regularity of the early time light curves.  As in \citet{Mummery+2024}, we discard TDEs that lack evidence for any distinct late-time emission component after fitting an exponential decay with flat plateau model (see \S \ref{sec:pheno}) to the light curve\footnote{More specifically, we discard 7 TDEs for which the late-time emission has a signal-to-noise ratio below 5 following comparison between a best-fit exponential model for the early time decay and a flat plateau fit \citep[see][for a more detailed description]{Mummery+2024}.}; this eliminates a further \red{7} \blue{5} TDEs from our sample, bringing us to our final sample of \red{38} \blue{40} TDEs.


\section{Plateau Fitting: Phenomenological}
\label{sec:pheno}

Late-time TDE plateau luminosities are often assumed to remain flat over time, or at least effectively flat over the time of observations \citep{Mummery+2024}. Nevertheless, spreading disk luminosities will clearly dim over a long enough baseline \citep{Cannizzo+1990}, and some theoretical studies predict that the late-time TDE luminosity should evolve on timescales as short as years \citep{Alush&Stone2025}. Such evolution has already been detected in the exceptionally well studied TDE ASASSN-14li \citep{Wen+2023}, but has not yet been studied at the population level.  In this section, we use the available late-time observations to examine whether there is statistically significant evidence for such evolution in our sample of late-time TDE disks.

In this section, we use phenomenological models that are not tied to specific physical assumptions about the nature of TDEs, only assuming that the spectrum can be described by a blackbody. Using these models, we fit all available optical and UV photometry simultaneously. The light curves are separated into the rising phase (before the peak), the early post-peak fast decay phase, and the late-time component:
\begin{equation}
    \nu L(t)=\nu L_{\rm rise}(t)+\nu L_{\rm early}(t)+\nu L_{\rm late}(t)
\end{equation}
where $L$ is the spectral luminosity density at frequency $\nu$ and time, $t$ ($t=0$ at the light curve maximum).

For the rising phase, we follow \cite{vanVelzen+2019} and model the early part of the light curve with a Gaussian rise:
\begin{equation}
    \nu L_{\rm rise}(t)=\nu L_{\rm peak}\frac{B_\nu(T_{\rm early})}{B_{\nu_0}(T_{\rm early})} e^{-\frac{(t-t_{\rm peak})^2}{2\sigma_{\rm rise}^2}}
\label{eq: rise model}
\end{equation}
for $t\leq t_{\rm peak}$ where $B_\nu$ is the Planck function at frequency $\nu$, and $\nu_0=10^{15}$Hz is a reference frequency.
The free parameters of the rising-phase model are the time of peak $t_{\rm peak}$, the luminosity at peak $L_{\rm peak}$, the temperature near peak $T_{\rm early}$, and the rise time $\sigma_{\rm rise}$. For TDEs in which the peak was not observed, $t_{\rm peak}$ is fixed to the time of the first observation, and this phase is not relevant for constraining the rest of the light curve parameters. 

For the early-time post-peak decaying component, we also follow \cite{vanVelzen+2019} and consider two different phenomenological models: first, an exponential decay,
\begin{equation}
    \nu L_{\rm early}(t)=\nu L_{\rm peak}\frac{B_\nu(T_{\rm early})}{B_{\nu_0}(T_{\rm early})}
        e^{-\frac{(t-t_{\rm peak})}{\tau_{\rm decay}}}
\label{eq: early model exp}
\end{equation}
and second, a power-law decay,
\begin{equation}
    \nu L_{\rm early}(t)=\nu L_{\rm peak}\frac{B_\nu(T_{\rm early})}{B_{\nu_0}(T_{\rm early})} 
        \left(\frac{t-t_{\rm peak}}{t_{0,\rm{decay}}}+1\right)^{-p_{\rm decay}}.
\label{eq: early model PL}
\end{equation}
While theoretical work sometimes assumes a power-law decay for early time emission due to the well-understood power-law evolution of the mass fallback rate \citep{Rees1988}, exponential decays often fit these light curves well \citep{Holoien+2014}.  Because large TDE samples contain some flares with early time decays better fit by exponentials, and others better fit by power-laws \citep{Yao+2023}, we remain agnostic and consider both possibilities.  Each of these parametrized models apply for $t>t_{\rm peak}$, where the free parameters are the exponential decay rate $\tau_{\rm decay}$ for the exponential model, and the decay timescale $t_{0,\rm{decay}}$ together with the power-law index $p_{\rm decay}$ for the power-law model.

For the late-time component, we consider three different models (each of which is only applicable for $t>t_{\rm peak}$). The first is a flat plateau, following \cite{Mummery+2024}:
\begin{equation}
    \nu L_{\rm late}(t)=\nu L_{\rm plat}\frac{B_\nu(T_{\rm plat})}{B_{\nu_0}(T_{\rm plat})},
\label{eq: late model flat}
\end{equation}
where the free parameters are the plateau luminosity $L_{\rm plat}$ and the plateau blackbody temperature $T_{\rm plat}$.

However, as we also wish to explore possible time evolution in the plateau, we test two alternative models, which we refer to as tilted plateau or ``cuesta'' models in brief\footnote{In geomorphology, a cuesta is a terrain feature resembling a tilted plateau, so we use the term for variables related to tilted or time-evolving plateau emission.}. \\
(i) An exponential decay model:
\begin{equation}
    \nu L_{\rm late}(t)=\nu L_{\rm plat}\frac{B_\nu(T_{\rm plat})}{B_{\nu_0}(T_{\rm plat})}
        e^{-\frac{(t-t_{\rm peak})}{\tau_{\rm cuesta}}}
\label{eq: late model exp}
\end{equation}
with free parameters $L_{\rm plat}$, $T_{\rm plat}$, and the exponential decay rate of the plateau, $\tau_{\rm cuesta}$; and \\
(ii) a power-law decay model:
\begin{equation}
    \nu L_{\rm late}(t)=\nu L_{\rm plat}\frac{B_\nu(T_{\rm plat})}{B_{\nu_0}(T_{\rm plat})}
        \left(\frac{t-t_{\rm peak}}{t_{0,{\rm cuesta}}}+1\right)^{-p_{\rm cuesta}}
\label{eq: late model PL}
\end{equation}
with free parameters $L_{\rm plat}$, $T_{\rm plat}$, the characteristic plateau decay time $t_{0,{\rm cuesta}}$, and the power-law index $p_{\rm cuesta}$. 

Finally, since the late-time observations may not necessarily show evidence for a plateau, we also test a model in which the post-peak light curve follows a single power-law decay:
\begin{align}
    \nu L_{\rm early}(t)+\nu L_{\rm late}(t)=& \nu L_{\rm peak}\frac{B_\nu(T_{\rm early})}{B_{\nu_0}(T_{\rm early})} \nonumber \\
      & \times  \left(\frac{t-t_{\rm peak}}{t_{0,\rm{decay}}}+1\right)^{-p_{\rm decay}}.
\label{eq: model one PL}
\end{align}
In total, we try 6 phenomenological models per TDE for the combined early and late emission in our observed sample.  We use a Markov Chain Monte Carlo (MCMC) method \citep{Foreman-Mackey+2013} to sample the posterior distributions of the model parameters. 
The models are fit to all (unbinned) data points with $t>-100$ days. 
The likelihood is assumed to be Gaussian and includes an additional variance term, which allows the data uncertainties to be scaled by a factor $f$ \citep{vanVelzen+2019b}.

\begin{figure}
    \centering
    \includegraphics[width=85mm]{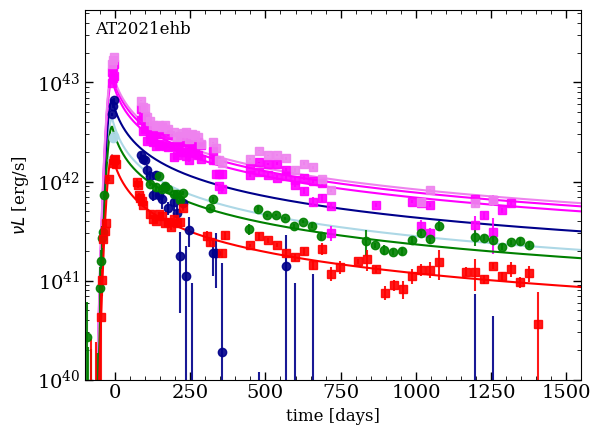}
    \includegraphics[width=85mm]{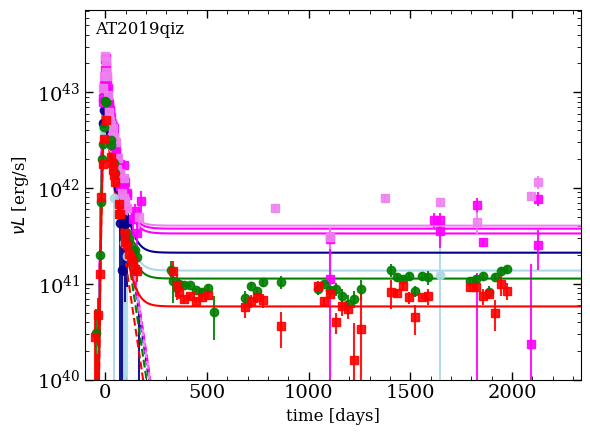}
    \includegraphics[width=85mm]{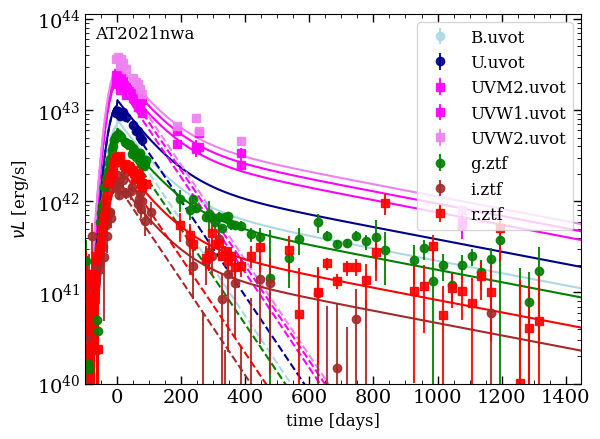}
    \caption{Three examples of observed multi-band light curves with the plateau phenomenological model with the lowest AIC: no plateau (single power-law; top), a flat plateau (middle), and a decaying plateau (bottom). The fitted models for early-time decay are shown as dashed lines. For presentation purposes, the data are binned in intervals of 3 days for $t<100$ days, 10 days for $t<365$ days, and 30 days for $t>365$ days.}
    \label{fig:light-curves}
\end{figure}

For all free parameters, we adopt \red{Gaussian priors with wide uncertainties, subject to the additional constraints: } \blue{uniform priors with bounds chosen to lie well outside the range of the final posterior distributions and are additionally subject to the following constraints: } $\sigma_{\rm rise}$, $\tau_{\rm decay}$, $\tau_{\rm cuesta}$, $t_{0,\rm{decay}}$, $t_{0,{\rm cuesta}}$, $p_{\rm decay}$, $p_{\rm cuesta}>0$ and $T_{\rm plat}>T_{\rm early}$. Three characteristic examples of the phenomenologically fitted light curves are shown in Fig. \ref{fig:light-curves}.

In this section, the SMBH masses are estimated either from the $M_\bullet$–$\sigma$ relation \citep{Greene+2020}:
\begin{equation}
    \log_{10}\left(\frac{M_\bullet}{M_\sun}\right)= 7.87+4.55\log_{10}\left(\frac{\sigma}{160\text{ km s}^{-1}}\right)
    \label{eq: scaling relation M-sigma}
\end{equation}
or from the host-galaxy mass scaling relation \citep{Greene+2020}:
\begin{equation}
    \log_{10}\left(\frac{M_\bullet}{M_\sun}\right)=7.43+1.61\log_{10}\left(\frac{M_{\rm gal}}{3\times10^{10}M_\odot}\right).
    \label{eq: scaling relation M-galaxy}
\end{equation}
The intrinsic scatter in the $M_\bullet$–$\sigma$ relation is $0.5$ dex, and in the host-galaxy mass relation it is $0.8$ dex.

\begin{figure}
    \centering
        \includegraphics[width=85mm]{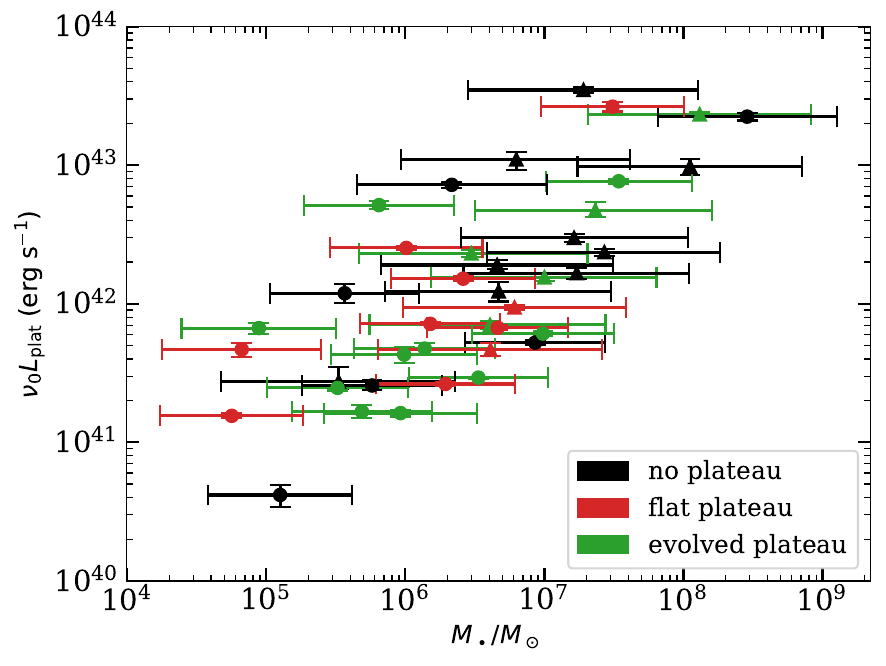}
        \includegraphics[width=85mm]{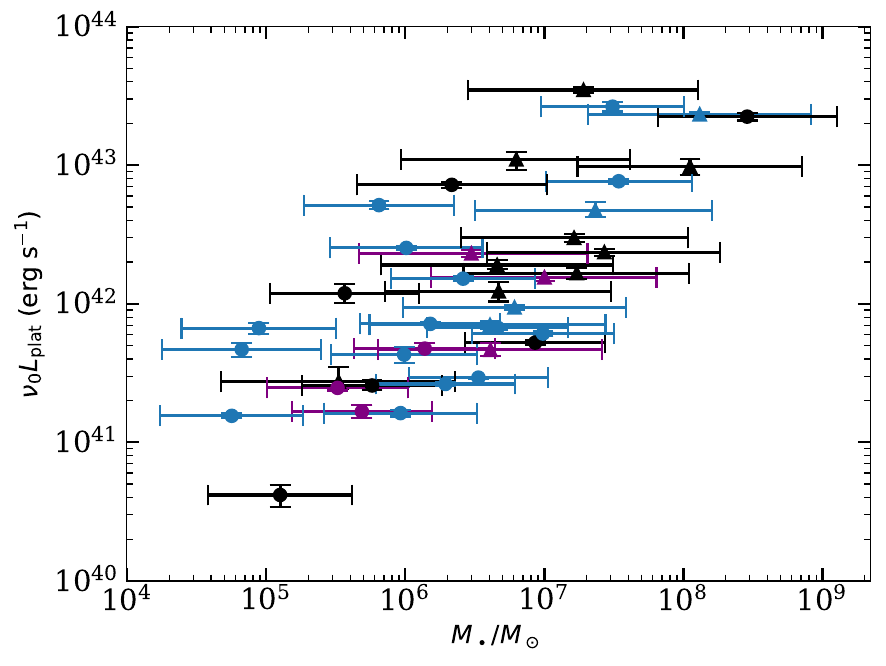}
    \caption{Characteristic plateau luminosities plotted against SMBH masses estimated from galaxy scaling relations ($M_\bullet$-$\sigma$ scaling relation in circles and $M_\bullet$-$M_{\rm gal}$ in triangles). The color coding indicates the qualitative nature of the best-fit (i.e. AIC minimizing) model. The top panel shows the nature of the plateau, while the bottom panel shows the early time. TDEs best fitted without a plateau (a single power-law decay) are shown in black; true (time-independent) plateaus are shown in red; and slowly evolving plateaus are shown in green.  Events whose best-fit models include a plateau with early-time exponential decay are shown in blue, while those with early-time power-law decay are shown in purple.
    }
    \label{fig:flat vs evolved}
\end{figure}

Our immediate goal is to estimate how many TDEs exhibit a flat late-time plateau compared to those that show time evolution, or alternatively, show no significant statistical evidence for a plateau at all. To this end, we use different combinations of early- and late-time models from Eqs. \eqref{eq: rise model}-(\ref{eq: model one PL}) and select the preferred model by minimize the Akaike Information Criterion (AIC). The fitted plateau luminosities are shown (as a function of the SMBH mass derived from scaling relations) in \cref{fig:flat vs evolved}, where the color of each point indicates the model with the lowest AIC. \blue{The preferred model for each TDE in our sample is summarized in \cref{tab:best model}. Representative examples of the fitted parameters for each model are presented in \cref{tab:one power law}, while the full dataset is provided in machine-readable format.}

We see that the plateau luminosity scales  with the SMBH mass, as previously found \citep{Mummery+2024}. However, for some TDEs, a flat, time-independent plateau is not the most favored model. In \cref{fig:flat vs evolved}, we classify the TDEs into three categories: 
\begin{enumerate}
    \item TDEs which are best fit by a single power law at both early and late times (i.e. Eq. \ref{eq: model one PL} minimizes the AIC), so that the best fit model is not straightforwardly interpretable as a plateau.
    \item TDEs with a flat plateau, where Eq. \ref{eq: late model flat} minimizes the AIC.
    \item TDEs with a cuesta (tilted, or evolving, plateau), where either Eq. \ref{eq: late model exp} or \ref{eq: late model PL} minimizes the AIC. 
\end{enumerate}
Representative examples of TDE light curves representing each category are shown in \cref{fig:light-curves}.

These categorizations are presented at the population level in Fig. \ref{fig:flat vs evolved}, where it is evident that only about one-third of the TDEs are best described by flat plateau models. Another third are better represented by a single power-law decay, indicating no evidence of a plateau, while the remaining third show plateaus that evolve over time. The early-time behavior is most often best fitted by an exponential decay, though some TDEs favor a power-law decay instead. Across these categories, there is little difference in the underlying SMBH mass distribution.  

Our fiducial categorization of late-time TDE emission simply uses the AIC to pick the best fitting model.  However, we have also compared the AIC difference between pairs of models ($\Delta$AIC) to test, in a more careful way, to what extent statistical evidence exists for the existence of plateau evolution, or even for the evolution of late-time plateaus in the first place.  First, we check for each TDE whether our ``no plateau'' model (Eq. \ref{eq: model one PL}) is ever disfavored at the level of $\Delta$AIC $\ge 10$ in comparison to any model with a late-time emission component.  In our sample, \red{14/38} \blue{16/40} TDEs lack this level of evidence against Eq. \ref{eq: model one PL}, and thus do not have strong evidence for a late time plateau.  Of the remaining 24 TDEs, we then check whether a perfectly flat plateau (Eq. \ref{eq: late model flat}) is ever disfavored at the level of $\Delta$AIC$\ge 10$ in comparison to other models with tilted late-time emission.  We conclude that \red{15/38} \blue{16/40} TDEs in our sample have evidence for a plateau {\it without clear evidence of time evolution}, while \red{9/38} \blue{8/40} TDEs show evidence for a tilted, time-evolving plateau.

\begin{figure}
    \centering
    \includegraphics[width=85mm]{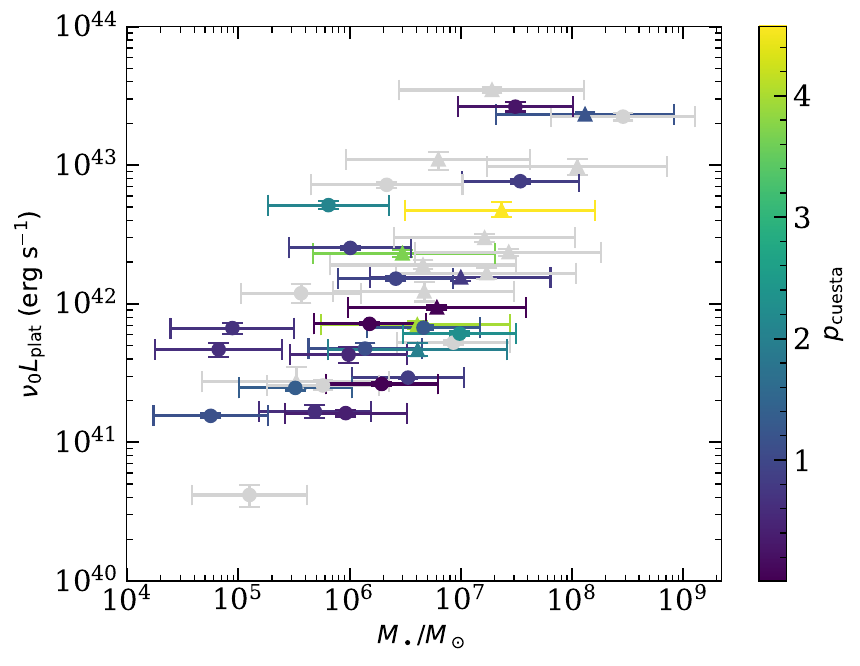}
    \caption{Characteristic plateau luminosities plotted against MBH masses estimated from galaxy scaling relations as in \cref{fig:flat vs evolved}.  Here all late-time plateau light curves have been fit with a power-law $L_{\rm late} \propto t^{-p_{\rm cuesta}}$ using \cref{eq: late model PL}, and the color coding represents the best-fit power law index $p_{\rm cuesta}$. The points in grey are those TDEs that show no statistical evidence for a plateau. 
    }
    \label{fig:lum MBH power law}
\end{figure}

To test how quickly or slowly the TDE plateaus evolve, we fit all TDEs that show evidence of a plateau with a late-time power-law decay model, as described in \cref{eq: late model PL}. The corresponding power-law indices range between \red{$0-2.6$} \blue{$0-4.6$} and are shown in \cref{fig:lum MBH power law}. From this figure, we see that some TDEs are best fitted with a steep (fast-decaying) power-law index. This suggests that greater care is needed when interpreting the late-time behavior of the plateau.

\section{Plateau Fitting: Disk Models}
\label{sec:models}

Now that we have examined the nature of each TDE's plateau emission (or lack thereof) in a phenomenological, theory-agnostic way, in this section we model the late-time plateau using a more realistic theoretical framework. Here we fit only the TDEs that show evidence for a plateau (either flat or tilted) based on the results from \S \ref{sec:pheno}.

\subsection{Magnetized Disk Model}

\label{subsec: mag disk}
 We use a 1D, time-dependent thin disk model for a highly magnetized accretion disk, based on the standard Shakura-Sunyaev model as modified in \cite{Alush&Stone2025}. In this model, gravity is Newtonian and the gas in the disk moves with a Keplerian angular frequency $\Omega_{\rm K}=\sqrt{GM_\bullet/R^3}$, where $G$ is the gravitational constant, $M_\bullet$ is the SMBH mass, and the relevant disk radii $R$ are assumed to be non-relativistic (i.e. large in comparison to the gravitational radius $r_{\rm g} = GM_\bullet / c^2$). The disk's surface density $\Sigma(R, t)$ evolves according to the diffusion equation: 
\begin{equation}
\frac{\partial\Sigma}{\partial t}=\frac{3}{R}\frac{\partial}{\partial R}\left\{R^{1/2}\frac{\partial}{\partial R}\left[\nu\Sigma R^{1/2}\right]\right\}    
\label{eq:diffusion eq}
\end{equation}
where $\nu$ is the effective viscosity, which we assume follows the $\alpha$-disk prescription $\nu=\alpha H c_{\rm s}$ \citep{ShakuraSunyaev1973}. The disk height $H=c_{\rm s}/\Omega_{\rm K}$ is determined from vertical hydrostatic equilibrium, $c_{\rm s}=\sqrt{P/\rho}$ is the sound speed, $\rho=\Sigma/H$ is the mid-plane density, and $P$ is the disk pressure. 

In this model, $P$ is assumed to be dominated by magnetic fields, such that $P\approx P_{\rm m}=B^2/8\pi$, where $B$ is the magnetic field strength. This type of ``magnetically elevated disk'' was first proposed by \citet{Begelman&Pringle2007}, who argued for a specific value of $P_{\rm m}$ that could be found from a saturation criterion for the magnetorotational instability (MRI) \citep{Pessah&Psaltis2005}. Applying this picture to a TDE disk, the MRI amplifies the weak toroidal seed magnetic field originating from the disrupted star until its growth rate is suppressed by magnetic tension, leading to saturation of the magnetic field strength \citep{Pessah&Psaltis2005, Begelman&Pringle2007}. The resulting magnetic pressure is 
\begin{equation}
    P_{\rm m}=v_{\rm K}\rho\sqrt{\frac{k_{\rm B}T}{\mu m_{\rm p}}}
    \label{eq: mag pressure}
\end{equation}
where $k_{\rm B}$ is the Boltzmann constant, $\mu=0.6$ is the mean molecular weight for Solar metallicity gas, $m_{\rm p}$ is the proton mass, and $T$ is the mid-plane temperature of the disk. While this specific saturation criterion has been challenged by subsequent analytic work \citep{Begelman&Armitage2023}, it is in reasonably good agreement with modern radiation-MHD simulations \citep{Jiang+2019, Mishra+2022, Zhang+2025} of accretion disks around black holes.  As the magnetically elevated hypothesis also stabilizes TDE disks against thermo-viscous instability \citep{Kaur+2023}, bringing simple models in line with observations, we continue (as in \citealt{Alush&Stone2025}) to use the \citet{Begelman&Pringle2007} saturation criterion for magnetic pressure in black hole accretion disks.

The disk temperature is determined by the balance between viscous heating and radiative cooling:
\begin{equation}
    \frac{4\sigma_{\rm SB}T^4}{3\kappa_{\rm{es}}\Sigma}=\frac{9}{8}\nu\Sigma\frac{GM_\bullet}{R^3}
\end{equation}
where $\sigma_{\rm SB}$ is the Stefan-Boltzmann constant and $\kappa_{\rm{es}}=0.34\,\rm{cm}^2\rm{g}^{-1}$ is the Thomson electron scattering opacity. \blue{In the inner regions of the disk, electron scattering dominates the opacity, and therefore we neglect absorption.}

The effective viscosity represents the transport of angular momentum driven by magnetized turbulence. Under the assumptions described above, it can be expressed as:
\begin{equation}
\begin{aligned}
    \nu&=\nu_{0,\rm{m}}R^{5/7}\Sigma^{2/7}\\
    \nu_{0,\rm{m}}&=\left[\frac{27}{32\sigma_{\rm SB}}\alpha^8\kappa_{\rm{es}}GM_\bullet\left(\frac{k_{\rm B}}{\mu m_{\rm p}}\right)^4\right]^{1/7}.
\end{aligned}
\label{eq:viscosity}
\end{equation}

The surface density diffusion equation (\cref{eq:diffusion eq}) is a non-linear equation due to the dependence of the viscosity on both $R$ and $\Sigma$, and it generally cannot be solved analytically. 
\blue{Accordingly, in this paper, we solve this equation numerically. We impose boundary conditions by setting the inner radius, $R_{\rm in}$, at the innermost stable circular orbit (ISCO), which is located at $6r_{\rm g}$ for a non-spinning SMBH, where $r_{\rm g}=GM_\bullet/c^2$ is the gravitational radius. The outer boundary, $R_{\rm out}\gg R_{\rm in}$, is chosen to be sufficiently large such that the surface density $\Sigma$ vanishes at radii smaller than $R_{\rm out}$ at all times. At both boundaries, we impose $\Sigma = 0$.
For the initial condition, we assume that half of the disrupted stellar mass, $m_\star/2$, has circularized into a Gaussian distribution centered at the circularization radius $r_{\rm c} = 2r_{\rm t}$. Here, $r_{\rm t} = r_\star \left(M_\bullet / m_\star\right)^{1/3}$ is the tidal disruption radius, and $r_\star$ is the stellar radius. }
\red{Therefore, in this work we use a self-similar solution of the form:
\begin{equation}
\begin{aligned}
    \Sigma_{\rm m,S}(t,R)&=\left(\frac{\nu_{0,\rm S}}{\nu_{0,\rm{m}}}\right)^{7/2}R^{-5/9}\left(\frac{3\nu_{0,\rm S}t}{4}\right)^{-35/36}\times\\
    &\times\left[1-\frac{1}{52}R^{13/9}\left(\frac{3\nu_{0,\rm S}t}{4}\right)^{-13/18}\right]^{7/2} 
\end{aligned}
\label{eq:similaritysol}
\end{equation}
where $\nu_{0,\rm S}$ depends on the Shakura-Sunyaev $\alpha$ parameter, the mass of the SMBH, $M_\bullet$, and the mass of the disrupted star $m_\star$. } 
\red{We use the best-fit model for $\nu_{0,\rm S}$ from \cite{Alush&Stone2025}:
\begin{equation}
    \nu_{0,\rm S}=1.91\times 10^{15}\alpha^{1.08}\left(\frac{m_\star}{M_\odot}\right)^{0.40}\left(\frac{M_\bullet}{M_\odot}\right)^{0.19}\left(1+\frac{M_\bullet}{M_{\rm Hills}}\right)^{-0.46}.
\end{equation}
}

\red{
Here $M_{\rm Hills}(m_\star)$ is the Hills mass \citep{Hills1975},which represents the maximum SMBH mass capable of disrupting a star of mass  $m_\star$. It is defined by the condition that the tidal disruption radius,
$r_t=R_\star\left(\frac{M_\bullet}{m_\star}\right)^{1/3}$, must be outside the innermost bound circular orbit (IBCO), where $R_\star$ is the stellar radius. }

To calculate the disk luminosity, we assume that the disk is optically thick \red{ and that the emission is isotropic} \blue{and emits locally as a blackbody, while accounting for the projection of the disk surface along, $i$, the line of sight.} \blue{To account for electron scattering in the photosphere, we include a color-correction factor $f_{\rm col}$, such that the observed color temperature is $T_{\rm col} = f_{\rm col} T_{\rm eff}$ where $f_{\rm col}\sim (\kappa_{\rm es}/\kappa_{\rm ff})^{1/8}$ \citep{Davis+2006}, $\kappa_{\rm es}$ and $\kappa_{\rm ff}$ are the electron scattering and the free-free opacity and $T_{\rm eff}=T\left(\frac{4}{3\kappa_{\rm es}\Sigma}\right)^{1/4}$ is the effective temperature.} Therefore, the spectral luminosity $L_\nu$ at a frequency $\nu$ is given by the \blue{corrected} Planck blackbody distribution, $B_\nu$:
\begin{equation}
    \nu L_{\rm mag}(t)=\red{4} \blue{8} \pi^2 \blue{\cos(i)}\int_{R_{\rm in}}^{R_{\rm out}} \nu \blue{f_{\rm col}^{-4}} B_\nu\left(\blue{f_{\rm col}} T_{\rm eff}\right)R dR.
\label{eq: mag light curves}
\end{equation}
\red{where the effective temperature is $T_{\rm eff}=T\left(\frac{4}{3\kappa_{\rm es}\Sigma}\right)^{1/4}$. The inner boundary, $R_{\rm in}$, corresponds to the innermost stable circular orbit (ISCO), which is located at $6r_{\rm g}$ for a non-spinning SMBH, where $r_{\rm g}=GM_\bullet/c^2$ is the gravitational radius. The outer boundary, $R_{\rm out}$, is set by the radius where the surface density vanishes; for the self-similar solution, it is given by $R_{\rm out}=52^{9/13}\left(\frac{3\nu_{0,\rm S}t}{4}\right)^{1/2}$.}

\subsection{Plateau Fitting}
\label{subsec: mag fitting}
Now that we have a disk model for the late-time TDE luminosity, we can fit the observed TDE light curves. For the late-time phase, where $t>t_{\rm peak}$, we use the magnetized disk model described in \cref{eq: mag light curves}. This model includes three free parameters: the SMBH mass $M_\bullet$, the Shakura-Sunyaev viscosity parameter $\alpha$, and the stellar mass $m_\star$ of the disrupted star\footnote{As the tidal radius does not formally enter our calculations, the stellar mass $m_\star$ can be viewed more accurately as twice the initial disk mass: for example, it could be less than the original stellar mass in the case of a partial disruption \citep{Bortolas+2023, Broggi+2024}, or if there has been substantial mass loss in circularization outflows \citep{Metzger&Stone2016, BonnerotLu2020, Steinberg&Stone2024}.}. 
\red{We use wide Gaussian priors for these parameters and impose additional priors requiring $M_\bullet<M_{\rm Hills}(m_\star)$, $-7<\log (\alpha)<3$, $-2<\log(m_\star/M_\odot)<1$. }  \blue{We adopt a prior on the stellar mass motivated by the Kroupa initial mass function, restricting the range to $-2<\log(m_\star/M_\odot)<1$. For the remaining parameters, we use uniform priors requiring $10^3 M_\odot<M_\bullet<M_{\rm Hills}(m_\star)$, $-7<\log (\alpha)<3$, and $0<\cos(i)<1$, where $M_{\rm Hills}(m_\star)$ denotes the Hills mass \citep{Hills1975}, defined as the maximum SMBH mass capable of disrupting a star of mass  $m_\star$. This Hills mass is determined by requiring that the tidal disruption radius, $r_{\rm t}$, lies outside the innermost bound circular orbit (IBCO). To prevent the disk model from fitting the early-time emission, we impose a prior requiring the disk luminosity to be still rising during the first month following disruption. }

\begin{figure}
    \centering
    \includegraphics[width=85mm]{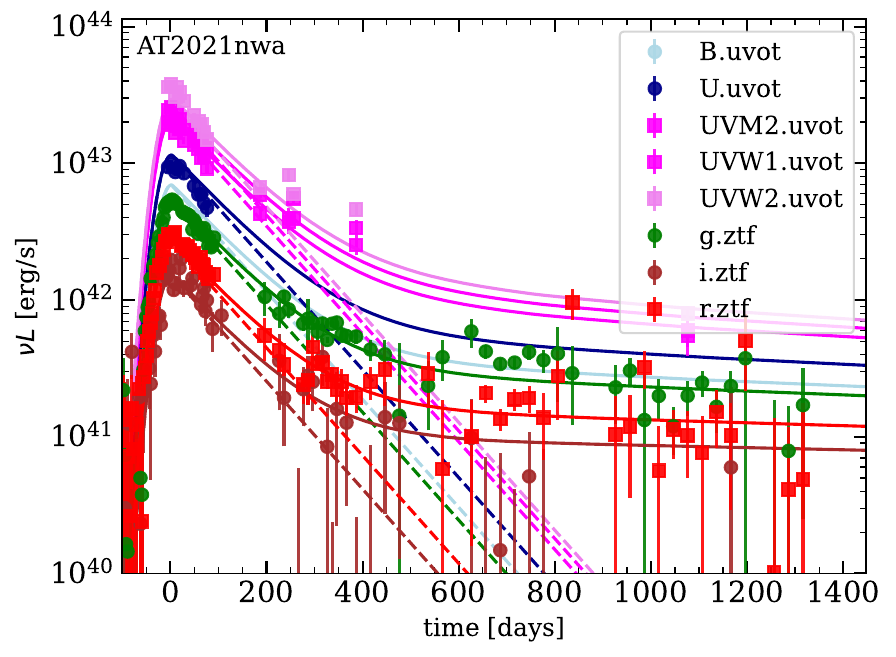}
    \caption{An example of observed multi-band light curves fitted using a the magnetized disk model. The early-time evolution is captured by an exponential decay (dashed lines), while the late-time emission is described by the magnetized disk model. For presentation purposes, the data are binned similar to \cref{fig:light-curves}. The complete figure set (25 images) is available in the online journal.}
    \label{fig:light-curves mag disk}
\end{figure}

For the early-time evolution, we use \cref{eq: rise model} to describe the rising phase before the peak. For the post-peak early-time phase, we use either the exponential decay model from \cref{eq: early model exp} or the power-law decay model from \cref{eq: early model PL}, depending on the best-fit case determined in \cref{fig:flat vs evolved}. The total luminosity is therefore given by
\begin{equation}
    \nu L(t)=\nu L_{\rm rise}(t)+\nu L_{\rm early}(t)+\nu L_{\rm mag}(t).
\end{equation}
\blue{An example of a TDE light-curve fit using the magnetized disk model is shown in \cref{fig:light-curves mag disk}, and the complete figure set is available in the online journal.} 
In this section, we include only the \red{26} \blue{25} TDEs for which the best-fit phenomenological model in \S \ref{sec:pheno} favors a plateau, either flat or evolving (cf. \cref{fig:flat vs evolved}). 

\blue{For the majority of the TDEs in our sample, the inclination with respect to the line of sight is poorly constrained, resulting in large uncertainties in $\cos(i)$. Incorporating inclination does not significantly affect the inferred values of $\alpha$ compared to the isotropic model. However, it does introduce additional scatter in the inferred SMBH masses.}

\begin{figure}
    \centering
    \includegraphics[width=85mm]{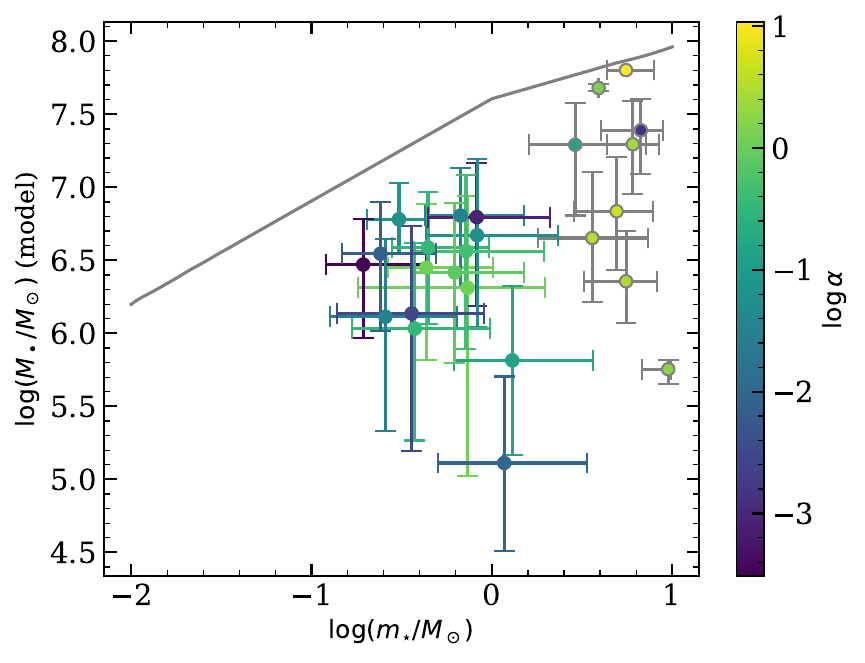}
    \caption{
    SMBH mass $M_\bullet$ as a function of the disrupted star’s mass $m_\star$, fitted using the magnetically elevated disk model. The color of each point indicates the Shakura–Sunyaev $\alpha$ parameter. The gray line marks the Hills mass. TDEs that cannot be explained by our model and are therefore excluded from the subsequent figures are shown with gray outlines. 
    }
    \label{fig:MBH vs mStar}
\end{figure}

To examine how well the magnetically elevated disk model fits the observed light curves, we show in \cref{fig:MBH vs mStar} the fitted SMBH mass as a function of the fitted stellar mass, along with the Hills mass boundary. We find that \red{10} \blue{9} TDEs run away to the Hills mass \blue{and/or the stellar mass} boundary, indicating that they cannot be explained within our model. This may suggest that the SMBHs in these events have high spins, which \blue{would enable disruption beyond the naive Hills mass included here.  Likewise, for SMBHs near the Hills mass, mass will be deposited quite close to the ISCO, worsening our neglect of inner-edge stresses and general relativity.} \red{are not included as free parameters in our model.} \red{In addition to this check on the Hills mass, we remove 2 more TDEs because our disk model is not a good description of the observations. In one case (AT2020mot), the late-time accretion disk model attempts to account for early-time luminosity, probably because the phenomenological early-time models do not capture this emission accurately. 
For another TDE (AT2021yzv), the early-time emission decays very slowly, and the fitted disk luminosity remains so subdominant throughout the light curve that the inferred disk parameters cannot be detected with meaningful significance. To conclude, we remove 12 TDEs because our model does not describe them accurately; in 10 of these cases, we suspect that the reason for this is the breakdown of our Newtonian treatment of accretion physics near the Hills mass.} These \red{12} \blue{9} TDEs are marked for completeness as light gray points in the subsequent analysis.

\figsetend

\begin{figure}
    \centering
    \includegraphics[width=85mm]{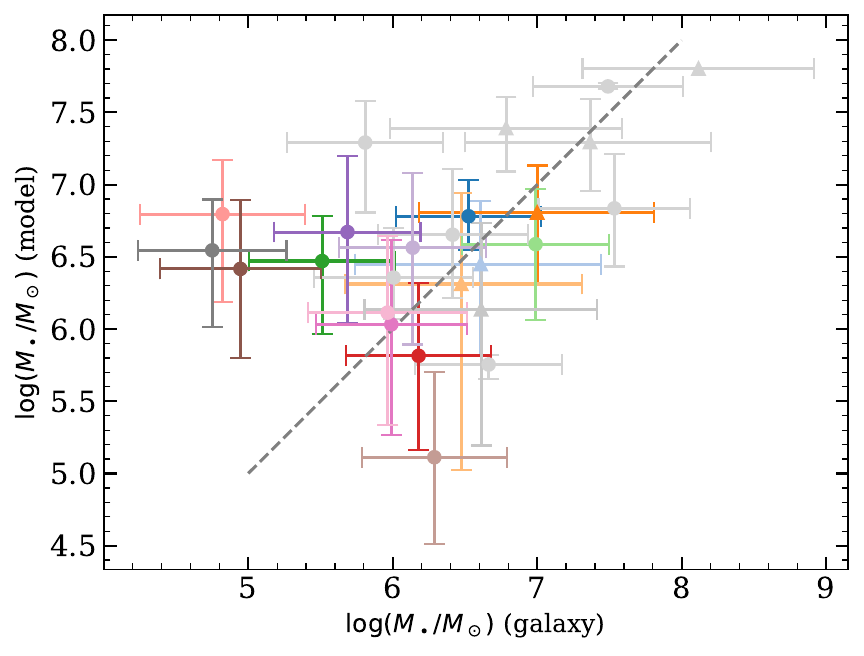}
    \caption{
    SMBH mass fitted from the magnetized disk model as a function of the SMBH mass from scaling relations (the $M_\bullet$-$\sigma$ scaling relation is shown as circles and $M_\bullet$-$M_{\rm gal}$ as triangles). The gray dashed line indicates where the SMBH masses are equal between the two approaches. The gray points are those TDEs  that are excluded because our disk model is not a correct description of the late-time observations, hence their $M_\bullet$ values (inferred from disk fitting) should not be trusted. 
    }
    \label{fig:mag disk fitted mass}
\end{figure}

In \cref{fig:mag disk fitted mass}, we show the SMBH masses fitted from the magnetized model as a function of those from the scaling relations (see Eqs. \ref{eq: scaling relation M-sigma}, \ref{eq: scaling relation M-galaxy}). However, without the excluded TDEs, our sample lacks the dynamic range to test any correlation between SMBH masses estimated from these two different approaches.

\begin{figure}
    \centering
    \includegraphics[width=85mm]{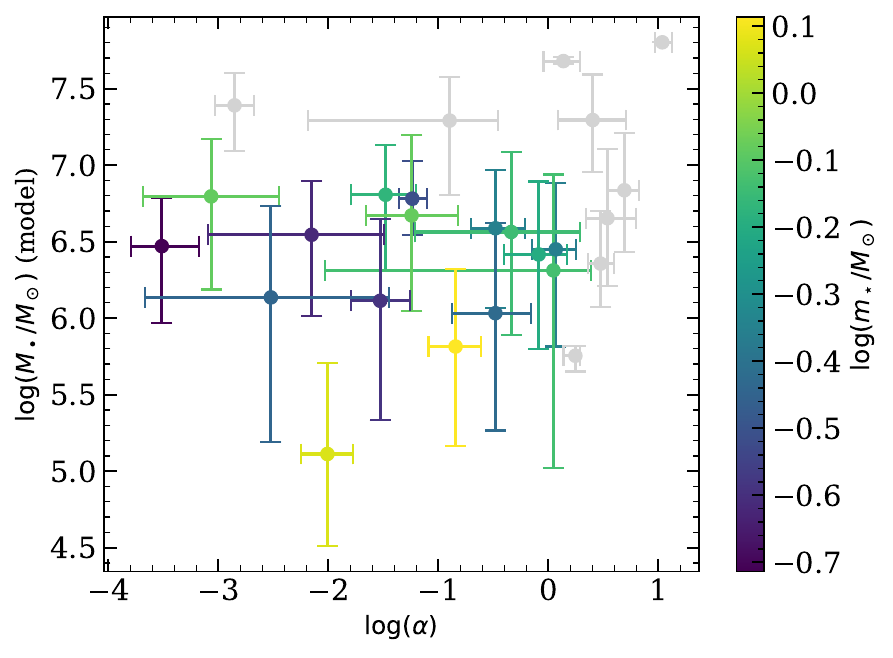}
    \caption{
    SMBH mass as a function of Shakura-Sunyaev $\alpha$-parameter fitted from the magnetized disk. The color of each point represents the stellar mass. The gray points are the TDEs  that are excluded because our disk model is not a correct description of the late-time observations, hence their $\alpha$ values should not be trusted.   The mean $\alpha$ among the well fitted TDEs is \red{$10^{-1.8}$} \blue{$10^{-1.5}$} with scatter of \red{$0.6$} \blue{$1$} dex.
    }
    \label{fig:mag disk M-alpha}
\end{figure}

A primary goal of this paper is to study the plateau evolution of late-time TDEs. Therefore, in \cref{fig:mag disk M-alpha}, we show the SMBH mass as a function of the Shakura-Sunyaev $\alpha$ parameter, which controls how quickly late time disks spread outwards. The prior on $\alpha$ was broad, yet the fitted values lie within the expected theoretical range: we find a mean \red{$\alpha=10^{-1.3}$ with scatter of 0.6 dex} \blue{$\alpha=10^{-1.5}$ with scatter of 1 dex}.  In the next section, we compare this range to values predicted by theoretical MHD simulations.

\section{Astrophysical Implications}
\label{sec:astro}

\subsection{Angular Momentum Transport}
\label{sec:alpha}

The $\alpha$ parameter is a dimensionless quantity that characterizes the efficiency of angular momentum transport by turbulence in the disk; larger values correspond to faster transport and thus a more rapid evolution of the plateau. Its physical origin remains uncertain, and different theoretical models or numerical simulations predict a wide range of possible values. Typical numerical estimates place $\alpha$ between $\sim 10^{-3}$ and $\sim 3 \times 10^{-1}$, with many studies clustering between $10^{-2}$ to $10^{-1}$ \citep{Hirose+2006, Davis+2010, Sorathia+2012, Penna+2013, Jiang+2019}; however, both lower and higher values can be produced under specific physical conditions.  In particular, while $\alpha<1$ is a strong theoretical expectation for local stresses, if accretion disks transport angular momentum via non-local global torques (e.g. magnetized winds or self-gravitating spiral arms), then larger effective $\alpha$ values are possible.

In \S\ref{subsec: mag fitting}, we examined a magnetically elevated disk model in which  the effective viscosity is an approximation to the magnetized turbulence powered by the MRI. Our fitted values \red{entirely} \blue{mostly} fall within the expected theoretical range: \red{$10^{-3} \lesssim \alpha \lesssim 0.4$} \blue{$-3.5 \lesssim \log(\alpha) \lesssim 0.07$}. This is a non-trivial result\footnote{\blue{However, we do note the presence of two outlier TDEs (AT2020vwl and AT2021yte) which have $\alpha < 0.01$ even within their $1\sigma$ error bars, in some tension with MHD simulations.}} because the prior we placed on $\alpha$ was very broad ($10^{-7} \le \alpha \le 10^3$).  In contrast, a subset of the TDEs that we discarded from our magnetized disk fits (because the MCMC ran away to the boundary of the priors) exhibit \blue{generally} higher $\alpha$ values. Our model should not be trusted for these cases, although if they had been found for well-converged MCMC posteriors, they could perhaps be explained by invoking additional sources of angular momentum transport, such as magnetized winds.

\blue{Some past work has also tried to estimate angular momentum transport efficiencies using observations of TDE disks.  For example, \citet{Guolo+25b} uses a linear viscosity model with a fixed viscous time, which can also be understood as a fixed prefactor to a linear $r-\phi$ stress component.  For two TDEs that are present in our sample, ASASSN-14li and AT2019dsg\footnote{\blue{Unfortunately, AT2019dsg was one TDE for which our model likely does not provide a good fit (i.e. the fitted SMBH mass ran away to the Hills mass boundary), so its results should be interpreted with caution.}}, they fit viscous timescales of $t_{\rm visc} = 470 \pm 30$ and $15 \pm 3$ days, respectively.  While we do not fit for viscous times directly, if we take our best-fit $\alpha$ values for these events ($\alpha = 10^{-1.2}$ and $\alpha = 10^{0.2} $, respectively), we find viscous times at $R = r_{\rm t}$ that, over the course of a decade, time-average to $t_{\rm visc}(r_{\rm t}) \approx 332$ days and $t_{\rm visc}(r_{\rm t}) \approx 18$ days, respectively.  It is interesting to see qualitative agreement between these two approaches despite (i) the difference in disk microphysics and (ii) the lack of X-ray information used in our fits, which provided an important additional source of information in \citet{Guolo+25b}.  We note that such qualitative agreement might not always emerge; in disks where angular momentum transport is governed by local stresses, local viscous times are likely strong functions of the disk aspect ratio $H/R$, as is assumed by the $\alpha$-ansatz.  While real accretion disks likely do not have constant stress-to-pressure ratios (i.e. do not satisfy a constant $\alpha$ assumption), the simplicity of late-time TDE disks may offer a way to test different microphysical prescriptions for average angular momentum transport efficiency.}

Aside from general questions about angular momentum transport in general accretion disks, constraints on $\alpha$ in TDE disks are particularly valuable for the study of TDEs and related phenomena.  The future wide-field UV survey satellite {\it ULTRASAT} is likely to discover thousands of TDEs at peak \citep{Shvartzvald+2024} and may even find dozens of TDE fossils \citep{Alush&Stone2025}: bare accretion disks discovered decades after peak light, with no observations of the early time light curve.  \citet{Alush&Stone2025} showed that if the photometric stability of the {\it ULTRASAT} camera does not exceed design sensitivity (5\%), then $\alpha$ will be a major uncertainty\footnote{However, if the photometric stability of the {\it ULTRASAT} camera achieves higher precision (1\%), then detection rates will be flux-limited rather than evolution-limited and $\alpha$ will have much less importance.} on fossil detection rates, with the number of detectable TDE fossils scaling almost linearly with $\alpha$.  

Angular momentum transport in TDE disks is also important for star-disk collision models \citep{Dai+2010} of quasi-periodic eruptions (QPEs).  Here, QPEs are thought to be triggered by pre-existing stars \citep{Linial&Metzger2023} or stellar mass compact objects \citep{Franchini+2023} crossing through transient TDE disks on inclined orbits.  Because $\alpha$ controls the rate of disk spreading, it determines the time of onset for QPEs.  It will also play a role in determining timing properties of QPEs in a subset of theoretical models, as we discuss in the next subsection.

\subsection{Disk Precession and Alignment}
\label{sec:tilt}

Because the stars that are tidally disrupted will typically approach the SMBH from a quasi-isotropic distribution of orbits, TDE disks will frequently circularize into initially tilted configurations \citep{Stone&Loeb2012}.  A tilted disk around a spinning SMBH will develop initially small warps due to differential nodal precession from the Lense-Thirring effect.  If the disk is geometrically thin, these linear warps propagate diffusively, and quickly grow into large-scale warps that drive alignment (the Bardeen-Petterson effect; \citealt{Bardeen&Petterson1975}).  By contrast, if the disk aspect ratio is sufficiently thick, linear warps propagate as bending waves that efficiently transport torques, and the disk experiences global, rigid-body nodal precession \citep{Papaloizou&Terquem1995}.  In $\alpha$-disk theory, the critical aspect ratio separating these two regimes is $H/R=\alpha$ \citep{Papaloizou&Lin1995}.  For realistic disks with significant Maxwell stresses (or Reynolds stresses from magnetized turbulence) the transition between these regimes is not as clear\footnote{Because the Bardeen-Petterson effect manifests only for thin disks, it is more challenging to simulate.  Recent GRMHD simulations have finally resolved the Bardeen-Petterson effect, though it appears to be substantially weaker than analytic predictions \citep{MoralesTeixeira+2014, Liska+2019}, and in some cases can be overwhelmed by disk tearing \citep{Liska+2021}.  See \citet{Fragile&Liska2024} for a recent review of tilted disk simulations.}, but MHD simulations confirm the expectation that tilted thick disks will precess as nearly solid bodies \citep{Fragile+2007, Liska+2018}.

As TDEs usually feature super-Eddington initial fallback rates, it is reasonable to expect that misaligned TDE disks will begin in the bending wave (rigidly precessing) regime \citep{Stone&Loeb2012, Franchini+2016}.  Their precession period $T_{\rm prec}$ will grow as they viscously spread outwards \citep{Stone+2013}.  Precession may have multiple observational manifestations, such as quasi-periodicity in hard X-ray/$\gamma$-ray emission from a precessing jet \citep{Stone&Loeb2012, Tchekhovskoy+2014, Franchini+2016}, quasi-periodicity in thermal soft X-rays from the inner disk \citep{Stone&Loeb2012, Franchini+2016, Wen+2020}, or even modulations in QPE timing properties \citep{Franchini+2023}.  There are hints of these signatures in observations of some TDEs \citep{Saxton+2012, Pasham+2024, Cao+2024}, though the observational evidence to date remains ambiguous.  Disk precession will eventually deactivate once hydrodynamic or magnetic torques align the disk into the Kerr equatorial plane, which will truncate the aforementioned quasi-periodicity while possibly creating late time radio signatures.  Precession may also effectively quench if disk spreading is so fast that $T_{\rm prec} \propto t^x$, with $x>1$ \citep{Stone+2013}. 

Past work on global precession of TDE disks usually assumes a standard Shakura-Sunyaev disk model.  The lack of observed thermal instability in late-time TDE disks that motivated our use of a magnetically elevated model will, in some cases, strongly affect precession rate and alignment time calculations.  In this subsection we will briefly revise standard precession and alignment calculations in the context of our magnetically elevated disk model, and then apply our observational constraints on $\alpha$ to estimate timescales for these phenomena in our sample of TDEs.

The rigid body precession timescale for a tilted TDE disk, $T_{\rm prec} = 2\pi / \Omega_{\rm prec}$, can be found from the rigid body precession frequency
\begin{equation}
    \Omega_{\rm prec} = \frac{\int_{R_{\rm in}}^{R_{\rm out}} \Sigma(R) \Omega(R) R^3 \Omega_\bullet(R) {\rm d}R}{\int_{R_{\rm in}}^{R_{\rm out}} \Sigma(R) \Omega(R) R^3{\rm d}R}, \label{eq:OmegaPrec}
\end{equation}
where $\Sigma(R)$ is the disk surface density as before, but unlike in \S \ref{subsec: mag disk},
\begin{equation}
    \Omega(R) = \frac{c}{R_{\rm g}}\left(\frac{R^{3/2}}{R_{\rm g}^{3/2}} +\chi_\bullet \right)^{-1}.
\end{equation}
This GR correction to the Newtonian orbital frequency has little impact on the fitting in \S \ref{subsec: mag fitting}, and so it is not used there, but it does have a larger effect on precession timescale calculations.  Eq. \ref{eq:OmegaPrec} also uses the local nodal precession frequency
\begin{equation}
    \Omega_\bullet(R) = \Omega(R)\left( \frac{2\chi_\bullet R_{\rm g}^{3/2}}{R^{3/2}} - \frac{3\chi_\bullet^2 R_{\rm g}^2}{2R^2} \right),
\end{equation}
accounting for both the Lense-Thirring effect (the leading order term, $\propto \chi_\bullet$) and the SMBH quadrupole moment (the $\propto \chi_\bullet^2$ term). 

\begin{figure}
    \centering
    \includegraphics[width=85mm]{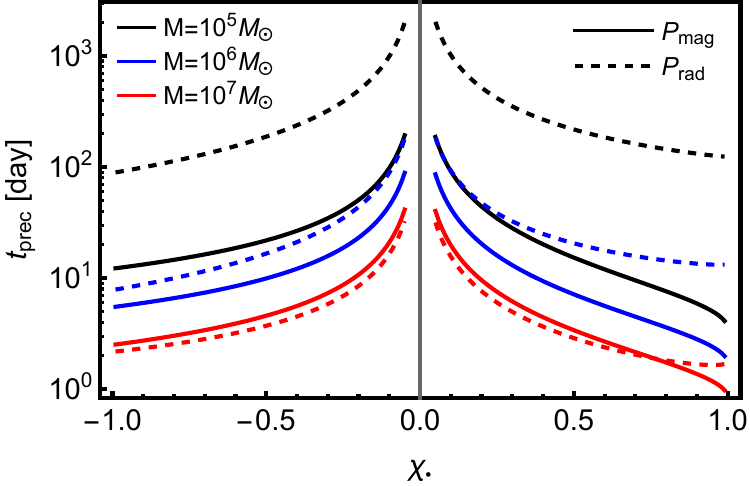}
    \caption{
     Precession time as a function of the SMBH spin for the steady-state solution of a Shakura-Sunyaev disk where the pressure dominant component is from radiation (dashed), and magnetic fields (solid). The SMBH masses are $10^5$ (black), $10^6$ (blue), $10^7$ (red). Shakura-Sunyaev $\alpha$-parameter is $\alpha=10^{-1.5}$. The mass accretion rate is $0.1M_{\rm edd}$, and the outer radius of the disk is fixed at $2r_{\rm c}$. For less massive SMBH, the precession is faster for a magnetized accretion disk. 
    }
    \label{fig:mag disk}
\end{figure}

We show rigid body disk precession timescales in Fig. \ref{fig:mag disk}, comparing steady state Shakura-Sunyaev models to steady state magnetically elevated models as in \citet{Kaur+2023}.  Here we set the disk inner radius $R_{\rm in}$ to be the ISSO (innermost stable spherical orbit; the tilted analogue of the ISCO) for a disk tilt of $\psi = 10^\circ$, and the disk outer radius $R_{\rm out} = 2r_{\rm c}$, where $r_{\rm c} = 2r_{\rm t}$, the circularization radius, is computed for the grazing disruption of a Solar-type star.  

In this figure, which represents an early time configuration of a TDE disk (before substantial viscous spreading has occurred), there is little difference between Shakura-Sunyaev and magnetically elevated precession times for $M_\bullet = 10^7 M_\odot$.  This is because $r_{\rm t}$ is quite close to the ISSO for such a large SMBH, and the disk resembles a narrow annulus for which the details of $\Sigma(R)$ do not have much impact on the integrals in Eq. \ref{eq:OmegaPrec}.  In contrast, the early time disks for $M_\bullet = 10^6 M_\odot$ and especially for $M_\bullet = 10^5 M_\odot$ have $R_{\rm out} \gg R_{\rm in}$ and therefore are quite sensitive to $\Sigma(R)$.  For these smaller SMBHs, magnetically elevated disk models typically see precession timescales $T_{\rm prec}$ that are 1-2 orders of magnitude {\it shorter} than unmagnetized ones.  The reason for this has to do with the surface density profile of the inner (radiation pressure dominated) region of a Shakura-Sunyaev disk, where $\Sigma$ increases with increasing $R$.  In contrast, a magnetically elevated disk has $\Sigma$ profiles that are relatively flat in $R$. Consequently, the magnetically elevated disks are capable of absorbing much more Lense-Thirring torque per unit (gas) angular momentum, and consequently see much shorter $T_{\rm prec}$.  This reduction in $T_{\rm prec}$ may be useful in explaining timing behavior seen in some QPEs \citep{Franchini+2023}.

\begin{figure}
    \centering
    \includegraphics[width=85mm]{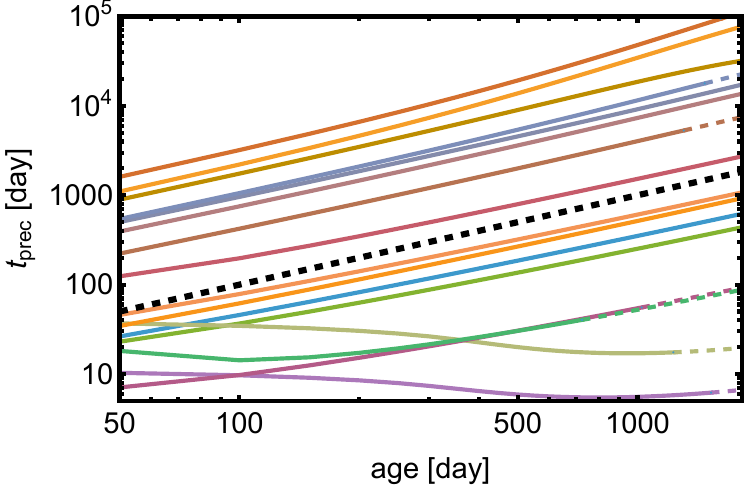}
    \caption{
     Precession times $T_{\rm prec}$ for each TDE in \S\ref{subsec: mag fitting} as a function of the TDE age $t$. The values of the SMBH mass, the Shakura-Sunyaev $\alpha$ parameter, and the stellar mass are taken from the fitted parameters of the magnetized disk model. The SMBH spin is assumed to be $\chi_\bullet=0.5$. The precession time is shown as a solid line for ages earlier than the last available observation, and as a dashed line for ages beyond the current observations.  The thick black dashed line shown for comparison is $T_{\rm prec}=t$; we see that at late times, most TDEs settle into a self-similar precession slowdown that appears to track this line.  Most TDE disks in our sample can therefore accumulate $\sim$ few-$10$ precession cycles.
    }
    \label{fig:precEvol}
\end{figure}

In reality, however, precession period will grow over time due to the outwards spreading of the disk \blue{\citep[see also][]{Teboul&Metzger2023}}.  We take our best-fit magnetized disk models from \S \ref{subsec: mag fitting} and use their fitted parameters ($M_\bullet$, $\alpha$, and $m_\star$) to estimate the time evolution of $T_{\rm prec}$ for \red{14} \blue{16} TDEs.  \red{Here we numerically solve the disk diffusion equation (Eq. \ref{eq:diffusion eq}) as in \citet{Alush&Stone2025}, rather than employing the self-similar solutions used for model fitting\footnote{While the self-similar solution matches numerical solutions well at large radii which dominate UV/optical emission, it is much less accurate at the small radii where Lense-Thirring torque is deposited \citep{Alush&Stone2025}.}.}  We show our results in Fig. \ref{fig:precEvol}, plotting the evolving $T_{\rm prec}$ against time $t$.  This figure shows that for almost all TDEs, precession times settle into a late-stage slowdown scaling as $T_{\rm prec} \propto t^x$, with $x \approx 1$, so that $T_{\rm prec} \approx y t$.  For \red{most} \blue{half of the} TDEs we have fit, $y < 1$ (with $y \sim 0.1-1$), indicating that precession of the disk will continue throughout its lifetime (absent alignment).  For \red{a minority of TDEs, $y \approx 2-3$} \blue{the remaining TDEs, $y \approx 2-30$}, indicating that precession quenches at early times. \blue{Some TDEs exhibit an initial decrease in the precession time before it rises again and settles into the $T_{\rm prec} \approx y t$ scaling. This behavior occurs in systems with lower values of $\alpha$, where the disk spreads more slowly. As a result, the disk has not yet reached the inner boundary, at the ISSO, so the angular momentum remains approximately constant and the precession time is dominated by $\Omega_\bullet$.}

This behavior is quite different from the precession evolution in thick spreading disks, where $T_{\rm prec} \propto t^{4/3}$ leads to quenching even absent alignment \citep{Stone+2013}.  We can gain a physical understanding of $T_{\rm prec}(t)$ by evaluating Eq. \ref{eq:OmegaPrec} at the order of magnitude level.  For most realistic disks, the upper integral (external torque) will be dominated by $R_{\rm in}$ and the lower integral (disk angular momentum) by $R_{\rm out}$.  Consequently, $\Omega_{\rm prec} \propto \Sigma(R_{\rm in}) \Omega(R_{\rm in}) R_{\rm in} / \Sigma(R_{\rm out}) \Omega(R_{\rm out}) R_{\rm out}^4$.  In the late-time, self-similar limit of a magnetically elevated spreading disk, $\Sigma(R, t) \propto R^{-5/9} t^{-35/36}$ \citep{Alush&Stone2025}, and $R_{\rm out} \propto t^{1/2}$.  Plugging these scalings in, we see that $T_{\rm prec}(t) \propto 1/\Omega_{\rm prec}(t) \propto t^{35/36}$, very close to the $x=1$ line drawn for reference in Fig. \ref{fig:precEvol}.

The disk nodal angle $\Phi$ will evolve as ${\rm d}\Phi/{\rm d}t = \Omega_{\rm prec}$.  In the limit of $T_{\rm prec} = yt$, the explicit evolution of the nodal angle will be $\Phi(t) = (2\pi / y) \ln(t/t_0)$, where $t_0$ is the time where global disk precession begins.  In other words, the accumulated number of precession cycles will be $N_{\rm prec} = y^{-1} \ln(t / t_0)$, which is $\sim$few$-10$ for most TDEs in Fig. \ref{fig:precEvol}.

Precession may also be cut short if the disk aligns into the Kerr midplane.  In the presence of very strong magnetic fields (as may be necessary to launch powerful jets), this alignment may occur quickly due to electromagnetic torques \citep{McKinney+2013, Polko&McKinney2017, Teboul&Metzger2023}.  The $B$-fields that produce a magnetically elevated state can be orders of magnitude weaker than those necessary to produce such a jet, however \citep{Kaur+2023}.

Alternatively, alignment may proceed hydrodynamically.  \citet{Franchini+2016} found that Bardeen-Petterson alignment is generally less important than alignment from internal torques produced by small twists and warps in the bending wave regime \citep{Bate+2000, Foucart&Lai2013}.  The hydrodynamic alignment time due to these internal torques will often be on the order of months to years \citep{Franchini+2016}.  We defer a detailed calculation of alignment processes in magnetically elevated disks to future work.

\section{Conclusions}
\label{sec:conclusions}

In this paper, we have revisited the phenomenon of late-time TDE disks by combining archival data with theoretical modeling.  Because late time TDE disks are generally dim and slowly evolving, past modeling has usually treated them as constant-luminosity plateaus.  While this approach has been fruitful, and has already led to the discovery of scaling relations that may aid parameter estimation in TDEs \citep{Mummery+2024}, it neglects the additional information encoded in late time light curve evolution.  We re-examined archival observations of a sample of 38 TDEs, fitting them in a theory-agnostic way that broke up their light curves into qualitatively discrete components.  We then applied physically motivated models for viscously spreading, magnetically elevated disks \citep{Alush&Stone2025} to re-fit the same set of data, extracting constraints on variables such as SMBH mass and the effective viscosity parameter $\alpha$. 

Our primary conclusions are as follows.
\begin{enumerate}
\item Out of the \red{38} \blue{40} TDEs we fit phenomenological models to, a large minority (\red{9} \blue{8} out of \red{38} \blue{40}) has strong statistical evidence ($\Delta$AIC$>10$) for time-evolving plateau emission.  

\item The magnetically elevated TDE disk model of \citet{Alush&Stone2025} achieves good fits for a majority the TDEs with the plateaus in our sample.  Although good fits cannot be achieved for a minority of \red{$12$} \blue{$9$} TDEs, we fit the remaining \red{$14$} \blue{16} and estimate SMBH mass $M_\bullet$, initial star/disk mass $m_\star$, and the effective Shakura-Sunyaev stress parameter $\alpha$.  The fitted $M_\bullet$ values are generally in agreement with masses estimated from galaxy scaling relations, while the $\alpha$ values \red{span a range of $10^{-3} < \alpha < 0.4$ with} \blue{have} a mean \red{$10^{-1.8}$} \blue{$10^{-1.5}$} with scatter of \red{$0.6$} \blue{1} dex, in reasonable agreement with GRMHD simulations of accretion physics. \blue{ Two outlier fits have best-fit $\alpha \sim 10^{-4} - 10^{-3}$, in tension with simulations.}
\item We have estimated global disk precession timescales $T_{\rm prec}$ for magnetically elevated TDE disks, and found that they may be orders of magnitude shorter than timescales for similar Shakura-Sunyaev disks.  We have estimated the time evolution of $T_{\rm prec}$ both theoretically ($T_{\rm prec} \propto t^{35/36}$) and in a data-driven way, using the best-fit values of $M_\bullet$, $m_\star$, and $\alpha$ from our sample.  We find that the typical late-time TDE disk experiences $\sim$few-10 precession cycles.
\end{enumerate}
Our modeling of TDE disks and their observable emission has been quite approximate in a number of ways that merit improvement in future work.  Despite our consideration of Lense-Thirring torques in \S \ref{sec:tilt}, we have treated the disk physics in a fundamentally Newtonian way, but in principle we could have used the fully relativistic generalization of Eq. \ref{eq:diffusion eq} \blue{\citep{Balbus2017, Mummery&Balbus2019}, as would be necessary to fold in X-ray information.} \red{(this would likely be a necessary step to fold in X-ray information, as in \citealt{Wen+2023}).}  Moving to a general relativistic (and spin-dependent) treatment of the ISCO may also be a necessary step in accurately fitting the \red{10} \blue{9} TDEs in our sample which could be fit phenomenologically, but for which our magnetically elevated disk models failed to converge to good fits (i.e. those TDEs which ran away to the Newtonian Hills mass).  Our phenomenological modeling of early-time TDE light curves was quite approximate, and improvements to this would better ``clean'' late-time TDE light curves of residual early-time components.  We have \red{neglected}\blue{found that} dust absorption and disk inclination with the observational line of sight \red{, both of which can alter the observed spectrum of TDE disks} \blue{have minimal impact on our fit results}.  Our disk models commit to a specific parametrization of magnetic pressure \citep{Begelman&Pringle2007} and neglect other pressure components; while we believe this choice is supported by the current generation of radiation-MHD accretion simulations (see \S \ref{subsec: mag disk}), it is ultimately an approximation that could be improved upon.

Most fundamentally, we have treated angular momentum transport via the classic Shakura-Sunyaev fudge factor, $\alpha$.  Real accretion disks feature non-trivial radial profiles of $\alpha$ \citep{Penna+2013, Jiang+2019}, so our $\alpha$ parameter can only be viewed as a spatio-temporally averaged value of the dimensionless stress in a TDE disk.  Nevertheless, we have seen that despite very broad priors (covering 10 orders of magnitude in $\alpha$, including possibly unphysical values with $\alpha > 1$), we recover posterior distributions for $\alpha$ \blue{that are broadly} within the range suggested by MHD simulations.  In principle, careful fitting of late time TDE observations may be applied to more detailed future models of angular momentum transport in slowly spreading accretion disks, constraining more fundamental aspects of the underlying MHD than the crude approximations of the $\alpha$ parameter.

As in \citet{Mummery+2024}, we have found a substantial minority of TDEs in our sample that lack strong evidence for a late time plateau.  Out of the \red{38} \blue{40} TDEs we fit phenomenologically, \red{14} \blue{16} lack strong evidence ($\Delta$AIC$<10$) against a single power-law model connecting early time decay and late time observations (a further \red{$12/14$} \blue{15/16} of these TDEs actually prefer the single power-law model as measured by the AIC). \blue{We do not find any correlation between the non-detectability of a plateau and the redshift of the source, its age, or the peak signal-to-noise ratio. As shown in \cref{fig:no plateau luminosity}, the late-time luminosities of the non-detected cases lies close to or above the scaling relation of \citet{Mummery+2024}.} This sub-sample calls into question the universality of the late-time plateau phase \blue{and may suggest that either (i) the disk has not yet formed, or (ii) the early-time component remains sufficiently bright to dominate the light curve.
} \red{though a more detailed investigation accounting for individual signal-to-noise of the late-time observations will be necessary to robustly probe this question.}

Late-time TDE disks were first considered decades ago as valuable test-beds for questions about the nature of accretion physics \citep{Cannizzo+1990}, and even the first samples of these UV sources \citep{vanVelzen+2019b} clearly contradicted simple predictions of $\alpha$-disk theory \citep{Shen&Matzner2014}.  While these disks may be usefully approximated as constant-luminosity plateaus for some applications \citep{Mummery+2024}, we have shown in this work that there is more information hiding in the slowly evolving light curves of TDEs years to decades post-peak.  This information may be useful for future parameter estimation in TDEs, but the basic setup of the problem is tantalizingly close to the oldest toy models in time-dependent accretion theory \citep{Pringle1981}.  Although future complexities may always emerge, this resemblance suggests that late-time TDE disks provide a natural laboratory for testing basic, unresolved questions in accretion theory, such as the nature of angular momentum transport.

\section*{Acknowledgements}
We acknowledge useful discussions with Xueyang Hu on aspects of light curve fitting. YA gratefully acknowledges the support of Azrieli fellowship.  YA and NCS gratefully acknowledge support from the Binational Science Foundation (grant No. 2020397) and the Israel Science Foundation (Individual Research Grant No. 2414/23). Support for this research was provided by the University of Wisconsin–Madison, Office of the Vice Chancellor for Research and Graduate Education, with funding from the Wisconsin Alumni Research Foundation. we
thank the referee for the constructive and helpful report. 

\bibliographystyle{aasjournal}
\bibliography{main}

\appendix

\section{Model fits}
\label{appendix:fits}
\cref{tab:best model} shows, for each TDE, the lowest AIC phenomenological model from \S\ref{sec:pheno}. In tables \cref{tab:one power law,tab:exp flat,tab:exp exp,tab:exp power-law,tab:power-law flat,tab:power-law exp}, we show a complete listing of fitted parameters for the phenomenological models described in \S\ref{sec:pheno}. The full set of fits for all phenomenological models described in \S\ref{sec:pheno}, as well as for the magnetized disk model described in \S\ref{sec:models}, is provided in machine-readable format.

\blue{A fraction of TDEs in our sample prefers the non-detected plateau model. We investigate whether this is connected to high redshift or short observational time, as suggested by previous work \citep{Mummery+2025b}. In contrast to that study, we define a ``non-detected plateau'' as a light curve that is best described by a single power law over both early and late times, compared to alternative models. We examine correlations between plateau detectability and redshift, SMBH mass, plateau luminosity, and the time of the last observation. While we find a correlation between plateau luminosity and redshift, we do not find a clear dependence of the non-detection fraction on either redshift or SMBH mass. All TDEs in our sample have observational baselines exceeding two years, and we do not find evidence that non-detections are associated with shorter observation durations. }

\begin{figure}
    \centering
    \includegraphics[width=85mm]{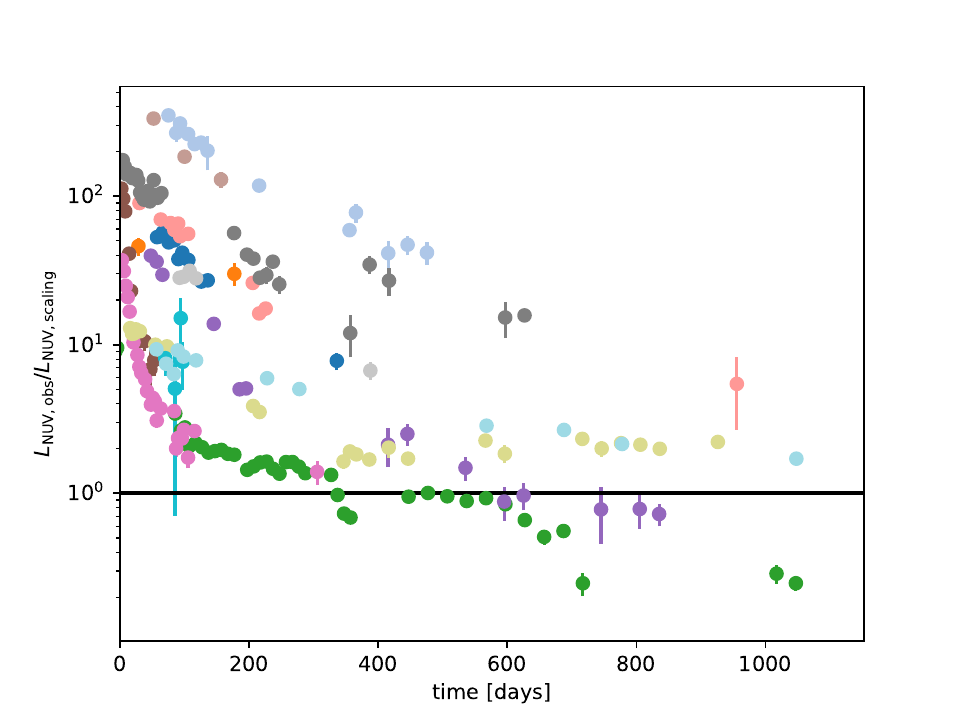}
    \includegraphics[width=85mm]{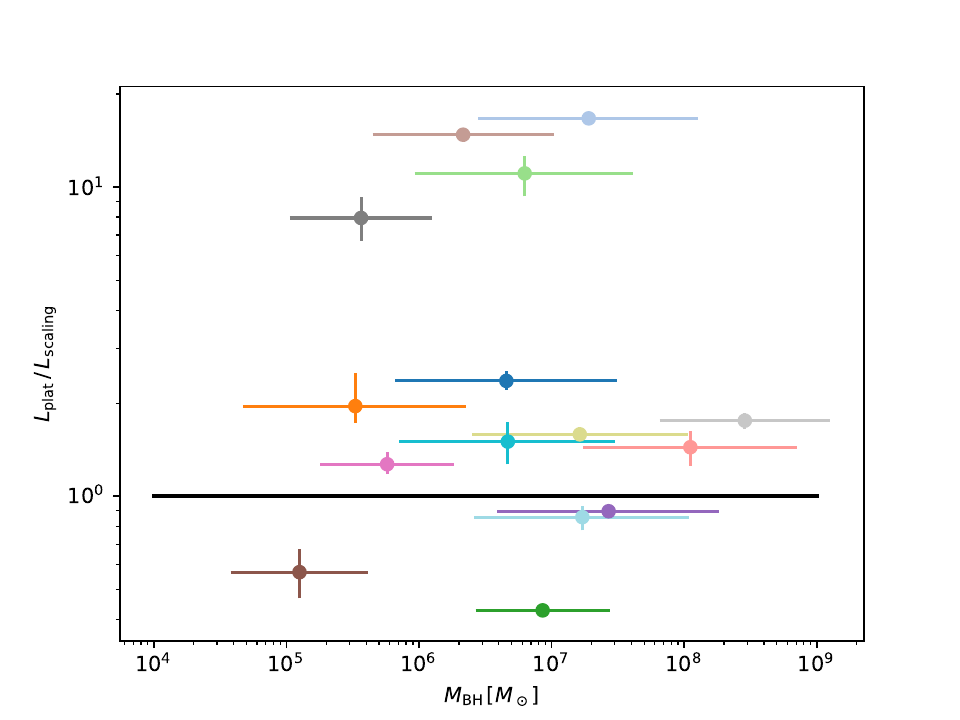}
    \caption{
        \blue{\textit{Left:} Observed NUV light curves of TDEs classified as non-plateau events, normalized by the NUV luminosity predicted by the scaling relation of \citep{Mummery+2024}.
        \textit{Right:} Plateau luminosities inferred from MCMC fits (exponential decay plus a constant plateau), normalized by the same scaling relation, shown as a function of black hole mass. Most non-plateau TDEs have luminosities that are close or above the predicted values.}
    }
    \label{fig:no plateau luminosity}
\end{figure}

\blue{Furthermore, in \cref{fig:no plateau luminosity}, most non-plateau events have observed plateau luminosities, as well as MCMC-inferred luminosities (from fits including an exponential decay plus a flat plateau), that are close or above the the values predicted by the scaling relation of \citep{Mummery+2024}. These results suggest that the absence of a detected plateau in some events is not solely due to observational limitations. Instead, it may reflect intrinsic differences in the emission evolution, such as cases where the accretion disk has not yet reached a sufficiently high luminosity to produce a detectable plateau \citep{Krolik+2025}, or where the early-time emission component remains sufficiently bright to dominate the light curve and cover an underlying plateau. To further test this, we suggest investigating the most luminous TDEs in this sample: AT2018jbv, AT2021axu, AT2021lo, and AT2018zr.}

\begin{table}[]
    \centering
    \begin{tabular}{|c|c|c|c|}
  
\hline 
TDE Name & early + late & TDE Name & early + late\\ 
 \hline 
 \hline 
ASASSN-14li & exponential + exponential & AT2018dyb & exponential + exponential\\
 \hline 
AT2018hco & one power-law & AT2018jbv & one power-law\\
 \hline 
AT2019ahk & power-law + exponential & AT2019dsg & exponential + falt\\
 \hline 
AT2019eve & one power-law & AT2020opy & power-law + exponential\\
 \hline 
AT2020vwl & power-law + exponential & AT2020ysg & exponential + exponential\\
 \hline 
AT2020yue & exponential + exponential & AT2021ehb & one power-law\\
 \hline 
AT2021lo & one power-law & AT2021nwa & exponential + exponential\\
 \hline 
AT2021uqv & exponential + falt & AT2021yzv & one power-law\\
 \hline 
AT2022dsb & exponential + falt & AT2022hvp & exponential + exponential\\
 \hline 
AT2023cvb & one power-law & AT2019azh & exponential + falt\\
 \hline 
AT2020mot & exponential + falt & AT2020wey & one power-law\\
 \hline 
AT2021axu & one power-law & AT2021yte & exponential + falt\\
 \hline 
ASASSN-14ae & power-law + exponential & iPTF-16fnl & one power-law\\
 \hline 
AT2018hyz & power-law + exponential & AT2018lna & exponential + exponential\\
 \hline 
AT2018lni & exponential + exponential & AT2018zr & one power-law\\
 \hline 
AT2019qiz & exponential + falt & AT2020qhs & one power-law\\
 \hline 
AT2020zso & exponential + exponential & ASASSN-15oi & exponential + exponential\\
 \hline 
AT2022upj & one power-law & AT2021gje & exponential + falt\\
 \hline 
AT2021jsg & one power-law & AT2022lri & exponential + falt\\
 \hline 
AT2023mhs & power-law + flat & AT2022gri & one power-law\\
 \hline

    \end{tabular}
    \caption{The model with the lowest AIC among the six models investigated in \S\ref{sec:pheno} for early and late times. }
    \label{tab:best model}
\end{table}

\begin{table}[]
    \centering

    \begin{tabular}{|c|c|c|c|c|c|}
      
\hline 
TDE Name & $\log(L_{\rm peak})$ &$\log(T_{\rm early})$ &$\log(t_0)$ &$p_{\rm decay}$ &AIC \\ 
 \hline 
 & $\log(\rm{erg}/\rm{s})$ &$\log(\rm{K})$ &$\log(s)$ &$1$ &\\ 
 \hline 
ASASSN-14li & $43.40_{-0.09}^{+0.11}$  &$4.70_{-0.04}^{+0.05}$  &$0.50_{-0.16}^{+0.13}$  &$0.78_{-0.01}^{+0.01}$  &$52530$\\
 \hline 
AT2018dyb & $43.66_{-0.02}^{+0.02}$  &$4.31_{-0.01}^{+0.01}$  &$1.46_{-0.04}^{+0.04}$  &$1.71_{-0.06}^{+0.06}$  &$7410$\\
 \hline 
AT2018hco & $44.08_{-0.04}^{+0.04}$  &$4.31_{-0.01}^{+0.01}$  &$1.46_{-0.07}^{+0.07}$  &$1.06_{-0.03}^{+0.03}$  &$179452$\\
 \hline 
AT2018jbv & $44.82_{-0.02}^{+0.02}$  &$4.31_{-0.01}^{+0.01}$  &$1.97_{-0.05}^{+0.05}$  &$1.16_{-0.05}^{+0.06}$  &$124465$\\
 \hline 
AT2019ahk & $43.94_{-0.02}^{+0.02}$  &$4.26_{-0.01}^{+0.01}$  &$1.03_{-0.05}^{+0.05}$  &$0.95_{-0.02}^{+0.02}$  &$17346$\\
 \hline 
AT2019dsg & $43.53_{-0.03}^{+0.04}$  &$4.25_{-0.01}^{+0.01}$  &$1.51_{-0.05}^{+0.05}$  &$1.07_{-0.04}^{+0.04}$  &$164890$\\
 \hline 
AT2019eve & $43.04_{-0.02}^{+0.02}$  &$4.00_{-0.01}^{+0.01}$  &$2.14_{-0.06}^{+0.06}$  &$1.84_{-0.12}^{+0.13}$  &$95990$\\
 \hline 
AT2020opy & $44.05_{-0.01}^{+0.01}$  &$4.20_{-0.01}^{+0.01}$  &$1.97_{-0.04}^{+0.04}$  &$1.63_{-0.06}^{+0.07}$  &$162114$\\
 \hline 
AT2020vwl & $43.86_{-0.09}^{+0.10}$  &$4.30_{-0.01}^{+0.01}$  &$1.06_{-0.10}^{+0.09}$  &$1.36_{-0.04}^{+0.04}$  &$120593$\\
 \hline 
AT2020ysg & $44.62_{-0.02}^{+0.02}$  &$4.26_{-0.01}^{+0.01}$  &$2.20_{-0.04}^{+0.04}$  &$1.60_{-0.07}^{+0.08}$  &$65512$\\
 \hline 
AT2020yue & $44.07_{-0.01}^{+0.01}$  &$3.98_{-0.00}^{+0.00}$  &$2.16_{-0.04}^{+0.04}$  &$1.94_{-0.10}^{+0.12}$  &$27952$\\
 \hline 
AT2021ehb & $42.93_{-0.03}^{+0.04}$  &$4.34_{-0.01}^{+0.01}$  &$1.42_{-0.07}^{+0.07}$  &$0.75_{-0.03}^{+0.03}$  &$109354$\\
 \hline 
AT2021lo & $44.12_{-0.09}^{+0.10}$  &$4.29_{-0.08}^{+0.11}$  &$1.61_{-0.15}^{+0.14}$  &$0.92_{-0.07}^{+0.08}$  &$53214$\\
 \hline 
AT2021nwa & $43.23_{-0.01}^{+0.01}$  &$4.47_{-0.01}^{+0.01}$  &$1.96_{-0.05}^{+0.05}$  &$1.49_{-0.07}^{+0.07}$  &$88271$\\
 \hline 
AT2021uqv & $43.58_{-0.01}^{+0.01}$  &$4.21_{-0.01}^{+0.01}$  &$1.32_{-0.05}^{+0.05}$  &$0.73_{-0.02}^{+0.02}$  &$198748$\\
 \hline 
AT2021yzv & $44.72_{-0.01}^{+0.01}$  &$4.31_{-0.01}^{+0.01}$  &$2.59_{-0.07}^{+0.07}$  &$3.32_{-0.34}^{+0.42}$  &$60964$\\
 \hline 
AT2022dsb & $43.13_{-0.05}^{+0.05}$  &$5.89_{-0.13}^{+0.08}$  &$0.09_{-0.07}^{+0.13}$  &$0.41_{-0.02}^{+0.02}$  &$33524$\\
 \hline 
AT2022hvp & $45.20_{-0.08}^{+0.10}$  &$4.70_{-0.06}^{+0.08}$  &$0.78_{-0.11}^{+0.10}$  &$1.23_{-0.04}^{+0.05}$  &$35448$\\
 \hline 
AT2023cvb & $44.24_{-0.05}^{+0.06}$  &$4.30_{-0.01}^{+0.01}$  &$1.75_{-0.08}^{+0.08}$  &$1.64_{-0.06}^{+0.07}$  &$62533$\\
 \hline 
AT2019azh & $44.05_{-0.02}^{+0.02}$  &$4.43_{-0.01}^{+0.01}$  &$0.01_{-0.01}^{+0.01}$  &$0.72_{-0.01}^{+0.01}$  &$213261$\\
 \hline 
AT2020mot & $43.64_{-0.01}^{+0.01}$  &$4.23_{-0.01}^{+0.01}$  &$1.54_{-0.03}^{+0.03}$  &$1.06_{-0.03}^{+0.03}$  &$153722$\\
 \hline 
AT2020wey & $42.83_{-0.02}^{+0.02}$  &$4.29_{-0.01}^{+0.02}$  &$0.87_{-0.07}^{+0.07}$  &$1.43_{-0.08}^{+0.09}$  &$99357$\\
 \hline 
AT2021axu & $44.40_{-0.01}^{+0.01}$  &$4.42_{-0.01}^{+0.01}$  &$1.86_{-0.03}^{+0.03}$  &$1.77_{-0.06}^{+0.06}$  &$154406$\\
 \hline 
AT2021yte & $43.40_{-0.05}^{+0.05}$  &$4.37_{-0.03}^{+0.03}$  &$1.05_{-0.14}^{+0.14}$  &$1.05_{-0.09}^{+0.10}$  &$53122$\\
 \hline 
ASASSN-14ae & $43.71_{-0.02}^{+0.03}$  &$4.25_{-0.01}^{+0.01}$  &$1.58_{-0.10}^{+0.08}$  &$2.41_{-0.26}^{+0.26}$  &$9966$\\
 \hline 
iPTF-16fnl & $42.81_{-0.03}^{+0.03}$  &$4.40_{-0.02}^{+0.02}$  &$0.04_{-0.03}^{+0.06}$  &$0.39_{-0.01}^{+0.01}$  &$17922$\\
 \hline 
AT2018hyz & $44.00_{-0.01}^{+0.01}$  &$4.21_{-0.00}^{+0.00}$  &$1.50_{-0.03}^{+0.03}$  &$1.39_{-0.04}^{+0.04}$  &$6825$\\
 \hline 
AT2018lna & $43.80_{-0.02}^{+0.02}$  &$4.48_{-0.02}^{+0.03}$  &$1.56_{-0.10}^{+0.11}$  &$1.55_{-0.13}^{+0.16}$  &$118675$\\
 \hline 
AT2018lni & $44.32_{-0.05}^{+0.04}$  &$5.41_{-0.45}^{+0.40}$  &$0.90_{-0.15}^{+0.14}$  &$0.81_{-0.05}^{+0.06}$  &$112986$\\
 \hline 
AT2018zr & $43.71_{-0.04}^{+0.04}$  &$4.13_{-0.01}^{+0.01}$  &$1.40_{-0.11}^{+0.11}$  &$0.90_{-0.05}^{+0.06}$  &$16753$\\
 \hline 
AT2019qiz & $42.90_{-0.02}^{+0.02}$  &$4.37_{-0.02}^{+0.02}$  &$0.01_{-0.01}^{+0.01}$  &$0.50_{-0.01}^{+0.01}$  &$140050$\\
 \hline 
AT2020qhs & $44.85_{-0.01}^{+0.01}$  &$4.30_{-0.01}^{+0.01}$  &$2.25_{-0.03}^{+0.04}$  &$1.97_{-0.08}^{+0.09}$  &$135085$\\
 \hline 
AT2020zso & $43.54_{-0.02}^{+0.02}$  &$4.26_{-0.01}^{+0.02}$  &$1.64_{-0.12}^{+0.12}$  &$2.19_{-0.26}^{+0.34}$  &$58729$\\
 \hline 
ASASSN-15oi & $44.37_{-0.05}^{+0.06}$  &$4.59_{-0.03}^{+0.03}$  &$1.32_{-0.06}^{+0.06}$  &$2.50_{-0.08}^{+0.09}$  &$22806$\\
 \hline 
AT2022upj & $43.30_{-0.03}^{+0.03}$  &$4.24_{-0.01}^{+0.01}$  &$1.64_{-0.16}^{+0.17}$  &$0.65_{-0.07}^{+0.09}$  &$40712$\\
 \hline 
AT2021gje & $44.49_{-0.02}^{+0.02}$  &$4.21_{-0.01}^{+0.01}$  &$1.34_{-0.06}^{+0.06}$  &$0.76_{-0.03}^{+0.03}$  &$190472$\\
 \hline 
AT2021jsg & $43.43_{-0.02}^{+0.02}$  &$4.16_{-0.01}^{+0.01}$  &$1.99_{-0.09}^{+0.10}$  &$2.43_{-0.34}^{+0.47}$  &$52388$\\
 \hline 
AT2022lri & $42.99_{-0.05}^{+0.06}$  &$4.40_{-0.01}^{+0.02}$  &$1.52_{-0.09}^{+0.08}$  &$1.22_{-0.05}^{+0.05}$  &$48544$\\
 \hline 
AT2023mhs & $44.13_{-0.05}^{+0.05}$  &$4.13_{-0.01}^{+0.01}$  &$1.63_{-0.07}^{+0.07}$  &$2.65_{-0.17}^{+0.19}$  &$12926$\\
 \hline 
AT2022gri & $43.11_{-0.01}^{+0.01}$  &$4.44_{-0.01}^{+0.01}$  &$2.92_{-0.07}^{+0.06}$  &$2.09_{-0.22}^{+0.20}$  &$28496$\\
 \hline 

    \end{tabular}
    \caption{Fits corresponding to the single power-law model described in \cref{eq: model one PL}. The quoted uncertainties represent $1\sigma$ errors. The reference frequency is fixed at $10^{15}$ Hz, and all logarithms are given in base 10. The complete dataset, including all fits and models, is provided in machine-readable format. }
    \label{tab:one power law}
\end{table}

\begin{table}[]
    \centering
    \begin{tabular}{|c|c|c|c|c|c|c|c|c|}
         
\hline 
TDE Name & $\log(\nu L_{\rm peak})$ &$\log(T_{\rm early})$ &$\log(\tau_{\rm decay})$ &$\log(\nu L_{\rm plat})$ &$\log(T_{\rm plat})$ &AIC \\ 
 \hline 
 & $\log(\rm{erg}/\rm{s})$ &$\log(\rm{K})$ &$\log(s)$ &$\log(\rm{erg}/\rm{s})$ &$\log(\rm{K})$ &\\ 
 \hline 
ASASSN-14li & $42.98_{-0.01}^{+0.01}$  &$4.53_{-0.02}^{+0.02}$  &$1.75_{-0.01}^{+0.01}$  &$41.47_{-0.01}^{+0.01}$  &$5.89_{-0.14}^{+0.08}$  &$52510$\\
 \hline 
AT2018dyb & $43.69_{-0.05}^{+0.05}$  &$4.29_{-0.02}^{+0.02}$  &$1.66_{-0.01}^{+0.01}$  &$41.85_{-0.03}^{+0.02}$  &$4.32_{-0.02}^{+0.02}$  &$7496$\\
 \hline 
AT2018hco & $43.59_{-0.01}^{+0.01}$  &$4.30_{-0.01}^{+0.01}$  &$2.35_{-0.01}^{+0.01}$  &$42.28_{-0.03}^{+0.03}$  &$4.36_{-0.04}^{+0.06}$  &$179545$\\
 \hline 
AT2018jbv & $44.81_{-0.03}^{+0.02}$  &$4.31_{-0.01}^{+0.01}$  &$2.24_{-0.02}^{+0.02}$  &$43.54_{-0.02}^{+0.02}$  &$4.32_{-0.01}^{+0.01}$  &$124506$\\
 \hline 
AT2019ahk & $43.75_{-0.01}^{+0.01}$  &$4.21_{-0.01}^{+0.01}$  &$1.88_{-0.01}^{+0.01}$  &$42.19_{-0.02}^{+0.02}$  &$4.41_{-0.03}^{+0.04}$  &$17370$\\
 \hline 
AT2019dsg & $43.42_{-0.04}^{+0.04}$  &$4.29_{-0.01}^{+0.01}$  &$2.23_{-0.01}^{+0.01}$  &$41.83_{-0.02}^{+0.02}$  &$4.30_{-0.01}^{+0.01}$  &$164868$\\
 \hline 
AT2019eve & $42.97_{-0.02}^{+0.02}$  &$4.00_{-0.01}^{+0.01}$  &$2.13_{-0.01}^{+0.01}$  &$41.44_{-0.05}^{+0.11}$  &$4.03_{-0.02}^{+0.07}$  &$96074$\\
 \hline 
AT2020opy & $43.94_{-0.01}^{+0.01}$  &$4.19_{-0.01}^{+0.01}$  &$2.11_{-0.01}^{+0.01}$  &$42.36_{-0.02}^{+0.02}$  &$4.19_{-0.01}^{+0.01}$  &$162087$\\
 \hline 
AT2020vwl & $43.16_{-0.02}^{+0.02}$  &$4.26_{-0.01}^{+0.01}$  &$1.96_{-0.01}^{+0.01}$  &$41.39_{-0.02}^{+0.02}$  &$4.50_{-0.04}^{+0.05}$  &$120623$\\
 \hline 
AT2020ysg & $44.63_{-0.03}^{+0.04}$  &$4.25_{-0.01}^{+0.01}$  &$2.16_{-0.01}^{+0.01}$  &$43.37_{-0.02}^{+0.02}$  &$4.26_{-0.01}^{+0.01}$  &$65543$\\
 \hline 
AT2020yue & $44.01_{-0.01}^{+0.01}$  &$3.98_{-0.01}^{+0.01}$  &$2.07_{-0.02}^{+0.02}$  &$42.67_{-0.05}^{+0.06}$  &$4.01_{-0.02}^{+0.03}$  &$28003$\\
 \hline 
AT2021ehb & $42.71_{-0.02}^{+0.02}$  &$4.33_{-0.01}^{+0.01}$  &$2.24_{-0.02}^{+0.02}$  &$41.72_{-0.02}^{+0.02}$  &$4.34_{-0.01}^{+0.01}$  &$109383$\\
 \hline 
AT2021lo & $43.71_{-0.13}^{+0.14}$  &$4.11_{-0.08}^{+0.11}$  &$2.28_{-0.05}^{+0.04}$  &$43.04_{-0.07}^{+0.06}$  &$5.17_{-0.50}^{+0.55}$  &$53220$\\
 \hline 
AT2021nwa & $43.17_{-0.01}^{+0.01}$  &$4.46_{-0.01}^{+0.01}$  &$2.08_{-0.01}^{+0.01}$  &$41.78_{-0.02}^{+0.02}$  &$4.47_{-0.01}^{+0.01}$  &$88345$\\
 \hline 
AT2021uqv & $43.37_{-0.01}^{+0.01}$  &$4.18_{-0.01}^{+0.01}$  &$2.12_{-0.01}^{+0.01}$  &$42.40_{-0.02}^{+0.02}$  &$4.19_{-0.01}^{+0.01}$  &$198695$\\
 \hline 
AT2021yzv & $44.69_{-0.01}^{+0.01}$  &$4.30_{-0.01}^{+0.01}$  &$2.20_{-0.01}^{+0.01}$  &$42.99_{-0.06}^{+0.05}$  &$4.32_{-0.02}^{+0.04}$  &$60979$\\
 \hline 
AT2022dsb & $43.14_{-0.06}^{+0.06}$  &$4.61_{-0.12}^{+0.20}$  &$1.29_{-0.04}^{+0.04}$  &$41.97_{-0.02}^{+0.02}$  &$5.90_{-0.13}^{+0.07}$  &$33480$\\
 \hline 
AT2022hvp & $44.93_{-0.03}^{+0.03}$  &$4.51_{-0.03}^{+0.03}$  &$1.46_{-0.02}^{+0.02}$  &$42.88_{-0.02}^{+0.02}$  &$4.52_{-0.03}^{+0.04}$  &$35426$\\
 \hline 
AT2023cvb & $43.88_{-0.02}^{+0.02}$  &$4.32_{-0.01}^{+0.01}$  &$2.09_{-0.02}^{+0.02}$  &$42.37_{-0.03}^{+0.02}$  &$4.33_{-0.01}^{+0.02}$  &$62546$\\
 \hline 
AT2019azh & $43.79_{-0.02}^{+0.02}$  &$4.30_{-0.01}^{+0.01}$  &$1.80_{-0.01}^{+0.01}$  &$41.85_{-0.01}^{+0.01}$  &$4.30_{-0.01}^{+0.01}$  &$212692$\\
 \hline 
AT2020mot & $43.54_{-0.01}^{+0.01}$  &$4.26_{-0.01}^{+0.01}$  &$2.06_{-0.02}^{+0.02}$  &$42.18_{-0.02}^{+0.02}$  &$4.27_{-0.01}^{+0.02}$  &$153675$\\
 \hline 
AT2020wey & $42.69_{-0.02}^{+0.03}$  &$4.31_{-0.02}^{+0.02}$  &$1.37_{-0.02}^{+0.02}$  &$40.62_{-0.08}^{+0.07}$  &$4.60_{-0.18}^{+0.44}$  &$99363$\\
 \hline 
AT2021axu & $44.30_{-0.01}^{+0.01}$  &$4.35_{-0.01}^{+0.01}$  &$1.84_{-0.01}^{+0.01}$  &$42.86_{-0.02}^{+0.02}$  &$5.69_{-0.33}^{+0.22}$  &$154495$\\
 \hline 
AT2021yte & $43.13_{-0.04}^{+0.04}$  &$4.22_{-0.02}^{+0.02}$  &$1.76_{-0.04}^{+0.04}$  &$41.67_{-0.05}^{+0.05}$  &$5.44_{-0.41}^{+0.38}$  &$53076$\\
 \hline 
ASASSN-14ae & $43.57_{-0.02}^{+0.02}$  &$4.26_{-0.02}^{+0.02}$  &$1.51_{-0.01}^{+0.01}$  &$41.22_{-0.04}^{+0.05}$  &$5.44_{-0.45}^{+0.38}$  &$9857$\\
 \hline 
iPTF-16fnl & $42.79_{-0.02}^{+0.02}$  &$4.32_{-0.02}^{+0.02}$  &$1.22_{-0.02}^{+0.02}$  &$41.41_{-0.03}^{+0.04}$  &$4.91_{-0.22}^{+0.51}$  &$17933$\\
 \hline 
AT2018hyz & $43.81_{-0.01}^{+0.01}$  &$4.21_{-0.01}^{+0.01}$  &$1.83_{-0.01}^{+0.01}$  &$41.68_{-0.04}^{+0.04}$  &$5.49_{-0.43}^{+0.35}$  &$6827$\\
 \hline 
AT2018lna & $43.67_{-0.02}^{+0.02}$  &$4.39_{-0.02}^{+0.02}$  &$1.84_{-0.03}^{+0.02}$  &$41.82_{-0.04}^{+0.04}$  &$4.43_{-0.03}^{+0.05}$  &$118642$\\
 \hline 
AT2018lni & $44.01_{-0.10}^{+0.12}$  &$4.44_{-0.14}^{+0.33}$  &$1.83_{-0.03}^{+0.03}$  &$42.71_{-0.03}^{+0.03}$  &$5.64_{-0.34}^{+0.25}$  &$112984$\\
 \hline 
AT2018zr & $43.49_{-0.02}^{+0.02}$  &$4.07_{-0.01}^{+0.01}$  &$2.10_{-0.02}^{+0.03}$  &$42.08_{-0.07}^{+0.07}$  &$4.65_{-0.17}^{+0.42}$  &$16807$\\
 \hline 
AT2019qiz & $43.15_{-0.01}^{+0.02}$  &$4.17_{-0.01}^{+0.01}$  &$1.48_{-0.01}^{+0.01}$  &$41.42_{-0.01}^{+0.01}$  &$4.33_{-0.02}^{+0.02}$  &$139613$\\
 \hline 
AT2020qhs & $44.74_{-0.01}^{+0.01}$  &$4.29_{-0.01}^{+0.01}$  &$2.17_{-0.01}^{+0.01}$  &$43.35_{-0.03}^{+0.03}$  &$4.30_{-0.01}^{+0.01}$  &$135135$\\
 \hline 
AT2020zso & $43.49_{-0.02}^{+0.02}$  &$4.24_{-0.02}^{+0.02}$  &$1.57_{-0.04}^{+0.04}$  &$41.63_{-0.06}^{+0.05}$  &$5.29_{-0.45}^{+0.47}$  &$58707$\\
 \hline 
ASASSN-15oi & $43.91_{-0.03}^{+0.02}$  &$4.54_{-0.04}^{+0.05}$  &$1.44_{-0.01}^{+0.01}$  &$41.21_{-0.02}^{+0.03}$  &$5.62_{-0.35}^{+0.27}$  &$22651$\\
 \hline 
AT2022upj & $43.20_{-0.02}^{+0.02}$  &$4.23_{-0.01}^{+0.01}$  &$2.19_{-0.05}^{+0.04}$  &$42.48_{-0.03}^{+0.03}$  &$4.24_{-0.01}^{+0.01}$  &$40713$\\
 \hline 
AT2021gje & $44.33_{-0.03}^{+0.02}$  &$4.17_{-0.02}^{+0.02}$  &$1.97_{-0.01}^{+0.01}$  &$43.42_{-0.03}^{+0.03}$  &$4.29_{-0.04}^{+0.04}$  &$190343$\\
 \hline 
AT2021jsg & $43.33_{-0.03}^{+0.03}$  &$4.10_{-0.02}^{+0.02}$  &$1.73_{-0.01}^{+0.01}$  &$42.09_{-0.07}^{+0.06}$  &$5.55_{-0.39}^{+0.31}$  &$52393$\\
 \hline 
AT2022lri & $42.66_{-0.02}^{+0.02}$  &$4.28_{-0.01}^{+0.01}$  &$2.06_{-0.02}^{+0.02}$  &$41.19_{-0.02}^{+0.02}$  &$5.71_{-0.30}^{+0.20}$  &$48497$\\
 \hline 
AT2023mhs & $43.93_{-0.03}^{+0.03}$  &$4.14_{-0.02}^{+0.02}$  &$1.54_{-0.03}^{+0.03}$  &$41.67_{-0.05}^{+0.05}$  &$4.27_{-0.05}^{+0.06}$  &$12926$\\
 \hline 
AT2022gri & $43.06_{-0.01}^{+0.01}$  &$4.44_{-0.01}^{+0.01}$  &$2.59_{-0.02}^{+0.02}$  &$42.22_{-0.04}^{+0.04}$  &$4.44_{-0.01}^{+0.01}$  &$28505$\\
 \hline 

    \end{tabular}
    \caption{Same as \cref{tab:one power law}, using the early-time exponential decay model (Eq. \ref{eq: early model exp}) and the flat plateau model (Eq. \ref{eq: late model flat}). }
    \label{tab:exp flat}
\end{table}

\begin{table}[]
    \centering
    \begin{tabular}{|c|c|c|c|c|c|c|c|c|}
    
\hline 
TDE Name & $\log(\nu L_{\rm peak})$ &$\log(T_{\rm early})$ &$\log(\tau_{\rm decay})$ &$\log(\nu L_{\rm plat})$ &$\log(T_{\rm plat})$ &$\log(\tau_{\rm cuesta})$ &AIC \\ 
 \hline 
 & $\log(\rm{erg}/\rm{s})$ &$\log(\rm{K})$ &$\log(s)$ &$\log(\rm{erg}/\rm{s})$ &$\log(\rm{K})$ &$\log(s)$ &\\ 
 \hline 
ASASSN-14li & $43.02_{-0.01}^{+0.01}$  &$4.51_{-0.02}^{+0.02}$  &$1.69_{-0.01}^{+0.01}$  &$41.64_{-0.01}^{+0.01}$  &$5.91_{-0.11}^{+0.06}$  &$3.37_{-0.02}^{+0.02}$  &$52286$\\
 \hline 
AT2018dyb & $42.26_{-0.03}^{+0.04}$  &$4.29_{-0.01}^{+0.01}$  &$2.61_{-0.04}^{+0.04}$  &$44.01_{-0.04}^{+0.04}$  &$4.32_{-0.01}^{+0.01}$  &$1.63_{-0.01}^{+0.01}$  &$7351$\\
 \hline 
AT2018hco & $43.78_{-0.02}^{+0.02}$  &$4.27_{-0.01}^{+0.01}$  &$1.99_{-0.02}^{+0.02}$  &$42.87_{-0.02}^{+0.03}$  &$4.30_{-0.02}^{+0.02}$  &$3.03_{-0.02}^{+0.02}$  &$179495$\\
 \hline 
AT2018jbv & $44.85_{-0.03}^{+0.02}$  &$4.32_{-0.01}^{+0.01}$  &$2.09_{-0.02}^{+0.02}$  &$43.97_{-0.04}^{+0.04}$  &$4.32_{-0.01}^{+0.01}$  &$3.14_{-0.05}^{+0.05}$  &$124468$\\
 \hline 
AT2019ahk & $43.79_{-0.01}^{+0.01}$  &$4.19_{-0.01}^{+0.01}$  &$1.82_{-0.01}^{+0.01}$  &$42.42_{-0.02}^{+0.02}$  &$4.38_{-0.02}^{+0.02}$  &$3.16_{-0.02}^{+0.03}$  &$17275$\\
 \hline 
AT2019dsg & $43.54_{-0.03}^{+0.04}$  &$4.26_{-0.01}^{+0.01}$  &$1.73_{-0.02}^{+0.02}$  &$42.53_{-0.02}^{+0.02}$  &$4.26_{-0.01}^{+0.01}$  &$2.91_{-0.02}^{+0.02}$  &$164987$\\
 \hline 
AT2019eve & $42.93_{-0.05}^{+0.03}$  &$3.99_{-0.02}^{+0.01}$  &$2.02_{-0.03}^{+0.02}$  &$42.16_{-0.10}^{+0.11}$  &$4.03_{-0.02}^{+0.08}$  &$2.81_{-0.06}^{+0.06}$  &$95993$\\
 \hline 
AT2020opy & $43.97_{-0.01}^{+0.01}$  &$4.20_{-0.01}^{+0.01}$  &$1.73_{-0.03}^{+0.03}$  &$43.41_{-0.03}^{+0.03}$  &$4.21_{-0.01}^{+0.01}$  &$2.55_{-0.02}^{+0.02}$  &$162167$\\
 \hline 
AT2020vwl & $43.36_{-0.04}^{+0.04}$  &$4.20_{-0.02}^{+0.02}$  &$1.69_{-0.03}^{+0.03}$  &$42.29_{-0.04}^{+0.04}$  &$4.40_{-0.02}^{+0.02}$  &$2.59_{-0.02}^{+0.03}$  &$120574$\\
 \hline 
AT2020ysg & $44.62_{-0.03}^{+0.04}$  &$4.25_{-0.01}^{+0.01}$  &$2.10_{-0.01}^{+0.01}$  &$43.71_{-0.05}^{+0.05}$  &$4.26_{-0.01}^{+0.01}$  &$3.07_{-0.06}^{+0.07}$  &$65497$\\
 \hline 
AT2020yue & $43.97_{-0.02}^{+0.02}$  &$3.97_{-0.01}^{+0.01}$  &$1.83_{-0.03}^{+0.03}$  &$43.52_{-0.05}^{+0.05}$  &$3.99_{-0.01}^{+0.01}$  &$2.52_{-0.03}^{+0.04}$  &$27949$\\
 \hline 
AT2021ehb & $42.82_{-0.03}^{+0.03}$  &$4.34_{-0.01}^{+0.01}$  &$1.86_{-0.03}^{+0.03}$  &$42.16_{-0.02}^{+0.02}$  &$4.35_{-0.01}^{+0.01}$  &$3.04_{-0.02}^{+0.03}$  &$109385$\\
 \hline 
AT2021lo & $43.70_{-0.19}^{+0.19}$  &$4.04_{-0.09}^{+0.11}$  &$1.83_{-0.08}^{+0.07}$  &$43.49_{-0.10}^{+0.10}$  &$4.47_{-0.15}^{+0.34}$  &$2.91_{-0.04}^{+0.06}$  &$53225$\\
 \hline 
AT2021nwa & $43.16_{-0.01}^{+0.01}$  &$4.45_{-0.01}^{+0.01}$  &$1.92_{-0.02}^{+0.02}$  &$42.39_{-0.04}^{+0.04}$  &$4.55_{-0.03}^{+0.03}$  &$2.81_{-0.03}^{+0.04}$  &$88257$\\
 \hline 
AT2021uqv & $43.43_{-0.02}^{+0.02}$  &$4.20_{-0.01}^{+0.01}$  &$1.83_{-0.03}^{+0.03}$  &$42.76_{-0.03}^{+0.03}$  &$4.22_{-0.01}^{+0.01}$  &$3.07_{-0.04}^{+0.04}$  &$198716$\\
 \hline 
AT2021yzv & $44.66_{-0.02}^{+0.01}$  &$4.30_{-0.01}^{+0.01}$  &$2.08_{-0.03}^{+0.03}$  &$43.93_{-0.13}^{+0.11}$  &$4.31_{-0.01}^{+0.01}$  &$2.65_{-0.06}^{+0.07}$  &$60965$\\
 \hline 
AT2022dsb & $43.14_{-0.06}^{+0.07}$  &$4.61_{-0.12}^{+0.19}$  &$1.29_{-0.04}^{+0.04}$  &$41.98_{-0.02}^{+0.02}$  &$5.90_{-0.14}^{+0.08}$  &$7.37_{-1.72}^{+1.74}$  &$33482$\\
 \hline 
AT2022hvp & $44.98_{-0.03}^{+0.03}$  &$4.50_{-0.03}^{+0.03}$  &$1.42_{-0.02}^{+0.02}$  &$43.09_{-0.03}^{+0.03}$  &$4.50_{-0.03}^{+0.03}$  &$2.98_{-0.05}^{+0.06}$  &$35387$\\
 \hline 
AT2023cvb & $44.03_{-0.03}^{+0.03}$  &$4.29_{-0.02}^{+0.01}$  &$1.82_{-0.03}^{+0.03}$  &$43.12_{-0.05}^{+0.04}$  &$4.30_{-0.01}^{+0.01}$  &$2.63_{-0.03}^{+0.03}$  &$62542$\\
 \hline 
AT2019azh & $43.79_{-0.02}^{+0.02}$  &$4.30_{-0.01}^{+0.01}$  &$1.80_{-0.01}^{+0.01}$  &$41.86_{-0.01}^{+0.01}$  &$4.30_{-0.01}^{+0.01}$  &$7.55_{-1.71}^{+1.66}$  &$212694$\\
 \hline 
AT2020mot & $43.59_{-0.01}^{+0.01}$  &$4.23_{-0.01}^{+0.01}$  &$1.81_{-0.01}^{+0.02}$  &$42.60_{-0.02}^{+0.02}$  &$4.24_{-0.01}^{+0.01}$  &$2.99_{-0.03}^{+0.03}$  &$153759$\\
 \hline 
AT2020wey & $42.81_{-0.02}^{+0.02}$  &$4.29_{-0.02}^{+0.02}$  &$0.85_{-0.05}^{+0.04}$  &$41.95_{-0.05}^{+0.05}$  &$4.32_{-0.02}^{+0.02}$  &$1.77_{-0.04}^{+0.04}$  &$99365$\\
 \hline 
AT2021axu & $43.92_{-0.02}^{+0.02}$  &$4.06_{-0.01}^{+0.01}$  &$1.71_{-0.01}^{+0.01}$  &$43.81_{-0.02}^{+0.02}$  &$5.87_{-0.16}^{+0.10}$  &$2.44_{-0.02}^{+0.02}$  &$154449$\\
 \hline 
AT2021yte & $43.13_{-0.05}^{+0.05}$  &$4.20_{-0.03}^{+0.03}$  &$1.69_{-0.12}^{+0.07}$  &$41.88_{-0.22}^{+0.19}$  &$5.54_{-0.40}^{+0.32}$  &$3.25_{-0.31}^{+3.44}$  &$53081$\\
 \hline 
ASASSN-14ae & $43.61_{-0.02}^{+0.02}$  &$4.24_{-0.02}^{+0.02}$  &$1.46_{-0.01}^{+0.01}$  &$41.56_{-0.05}^{+0.05}$  &$5.54_{-0.41}^{+0.32}$  &$3.28_{-0.09}^{+0.09}$  &$9858$\\
 \hline 
iPTF-16fnl & $42.79_{-0.02}^{+0.02}$  &$4.32_{-0.02}^{+0.02}$  &$1.22_{-0.02}^{+0.02}$  &$41.41_{-0.03}^{+0.04}$  &$4.87_{-0.20}^{+0.44}$  &$7.75_{-1.58}^{+1.48}$  &$17935$\\
 \hline 
AT2018hyz & $43.85_{-0.01}^{+0.01}$  &$4.18_{-0.01}^{+0.01}$  &$1.73_{-0.02}^{+0.02}$  &$42.48_{-0.11}^{+0.09}$  &$4.55_{-0.11}^{+0.25}$  &$2.80_{-0.03}^{+0.04}$  &$6802$\\
 \hline 
AT2018lna & $43.71_{-0.02}^{+0.02}$  &$4.42_{-0.02}^{+0.02}$  &$1.73_{-0.03}^{+0.04}$  &$42.15_{-0.08}^{+0.07}$  &$4.46_{-0.03}^{+0.05}$  &$3.19_{-0.11}^{+0.14}$  &$118640$\\
 \hline 
AT2018lni & $44.00_{-0.10}^{+0.11}$  &$4.41_{-0.13}^{+0.26}$  &$1.78_{-0.04}^{+0.04}$  &$42.86_{-0.07}^{+0.06}$  &$5.66_{-0.36}^{+0.24}$  &$3.52_{-0.14}^{+0.24}$  &$112976$\\
 \hline 
AT2018zr & $43.52_{-0.02}^{+0.02}$  &$4.00_{-0.01}^{+0.01}$  &$1.69_{-0.04}^{+0.04}$  &$42.87_{-0.04}^{+0.03}$  &$4.39_{-0.03}^{+0.04}$  &$2.63_{-0.02}^{+0.03}$  &$16781$\\
 \hline 
AT2019qiz & $43.15_{-0.02}^{+0.02}$  &$4.17_{-0.01}^{+0.01}$  &$1.48_{-0.01}^{+0.01}$  &$41.42_{-0.01}^{+0.01}$  &$4.33_{-0.02}^{+0.02}$  &$7.83_{-1.59}^{+1.50}$  &$139615$\\
 \hline 
AT2020qhs & $44.75_{-0.01}^{+0.01}$  &$4.30_{-0.01}^{+0.01}$  &$1.98_{-0.02}^{+0.02}$  &$44.07_{-0.05}^{+0.05}$  &$4.31_{-0.01}^{+0.01}$  &$2.75_{-0.04}^{+0.04}$  &$135089$\\
 \hline 
AT2020zso & $43.50_{-0.02}^{+0.02}$  &$4.24_{-0.02}^{+0.02}$  &$1.53_{-0.05}^{+0.05}$  &$41.80_{-0.12}^{+0.12}$  &$5.06_{-0.42}^{+0.58}$  &$3.27_{-0.27}^{+0.74}$  &$58706$\\
 \hline 
ASASSN-15oi & $43.93_{-0.02}^{+0.02}$  &$4.54_{-0.04}^{+0.05}$  &$1.43_{-0.01}^{+0.01}$  &$41.37_{-0.03}^{+0.04}$  &$5.62_{-0.37}^{+0.27}$  &$3.45_{-0.08}^{+0.09}$  &$22635$\\
 \hline 
AT2022upj & $43.20_{-0.02}^{+0.02}$  &$4.23_{-0.01}^{+0.01}$  &$2.17_{-0.08}^{+0.05}$  &$42.49_{-0.03}^{+0.11}$  &$4.24_{-0.01}^{+0.01}$  &$5.80_{-2.33}^{+2.75}$  &$40716$\\
 \hline 
AT2021gje & $44.32_{-0.03}^{+0.02}$  &$4.17_{-0.02}^{+0.02}$  &$1.97_{-0.01}^{+0.01}$  &$43.43_{-0.04}^{+0.04}$  &$4.30_{-0.04}^{+0.04}$  &$6.47_{-2.19}^{+2.40}$  &$190347$\\
 \hline 
AT2021jsg & $43.25_{-0.05}^{+0.04}$  &$4.05_{-0.02}^{+0.02}$  &$1.67_{-0.06}^{+0.05}$  &$42.66_{-0.32}^{+0.19}$  &$4.46_{-0.14}^{+0.81}$  &$2.43_{-0.17}^{+0.47}$  &$52394$\\
 \hline 
AT2022lri & $42.68_{-0.02}^{+0.02}$  &$4.27_{-0.01}^{+0.01}$  &$1.99_{-0.03}^{+0.03}$  &$41.45_{-0.08}^{+0.07}$  &$5.62_{-0.34}^{+0.27}$  &$3.09_{-0.10}^{+0.16}$  &$48515$\\
 \hline 
AT2023mhs & $43.99_{-0.03}^{+0.03}$  &$4.11_{-0.02}^{+0.02}$  &$1.45_{-0.03}^{+0.03}$  &$42.47_{-0.16}^{+0.18}$  &$4.24_{-0.06}^{+0.06}$  &$2.29_{-0.10}^{+0.11}$  &$12916$\\
 \hline 
AT2022gri & $43.04_{-0.02}^{+0.02}$  &$4.44_{-0.01}^{+0.01}$  &$2.54_{-0.04}^{+0.04}$  &$42.50_{-0.20}^{+0.12}$  &$4.44_{-0.01}^{+0.01}$  &$3.33_{-0.16}^{+0.52}$  &$28504$\\
 \hline

    \end{tabular}
    \caption{Same as \cref{tab:one power law}, but using the early-time exponential decay model  (Eq. \ref{eq: early model exp}) and the exponential plateau model (Eq. \ref{eq: late model exp}).}
    \label{tab:exp exp}
\end{table}

\begin{table}[]
    \centering
    \begin{tabular}{|c|c|c|c|c|c|c|c|c|}
     
  \hline 
TDE Name & $\log(\nu L_{\rm peak})$ &$\log(T_{\rm early})$ &$\log(\tau_{\rm decay})$ &$\log(\nu L_{\rm plat})$ &$\log(T_{\rm plat})$ &$\log(t_{0,\rm cuesta})$ &$p_{\rm cuesta}$ &AIC \\ 
 \hline 
 & $\log(\rm{erg}/\rm{s})$ &$\log(\rm{K})$ &$\log(s)$ &$\log(\rm{erg}/\rm{s})$ &$\log(\rm{K})$ &$\log(s)$ & 1 &\\ 
 \hline 
ASASSN-14li & $43.02_{-0.01}^{+0.01}$  &$4.51_{-0.02}^{+0.02}$  &$1.68_{-0.01}^{+0.01}$  &$41.71_{-0.01}^{+0.01}$  &$5.92_{-0.12}^{+0.06}$  &$2.95_{-0.08}^{+0.04}$  &$0.81_{-0.07}^{+0.05}$  &$52308$\\
 \hline 
AT2018dyb & $43.75_{-0.03}^{+0.03}$  &$4.30_{-0.01}^{+0.01}$  &$1.56_{-0.01}^{+0.01}$  &$42.42_{-0.03}^{+0.03}$  &$4.32_{-0.01}^{+0.01}$  &$2.96_{-0.06}^{+0.03}$  &$3.96_{-0.36}^{+0.32}$  &$7372$\\
 \hline 
AT2018hco & $43.72_{-0.08}^{+0.05}$  &$4.27_{-0.03}^{+0.02}$  &$1.85_{-0.04}^{+0.04}$  &$43.43_{-0.17}^{+0.19}$  &$4.30_{-0.01}^{+0.02}$  &$2.22_{-0.28}^{+0.30}$  &$1.17_{-0.10}^{+0.18}$  &$179463$\\
 \hline 
AT2018jbv & $44.81_{-0.04}^{+0.04}$  &$4.31_{-0.01}^{+0.01}$  &$2.05_{-0.03}^{+0.03}$  &$44.26_{-0.11}^{+0.16}$  &$4.32_{-0.01}^{+0.01}$  &$2.56_{-0.36}^{+0.31}$  &$1.10_{-0.21}^{+0.31}$  &$124467$\\
 \hline 
AT2019ahk & $43.78_{-0.02}^{+0.01}$  &$4.19_{-0.01}^{+0.01}$  &$1.79_{-0.01}^{+0.01}$  &$42.73_{-0.12}^{+0.31}$  &$4.36_{-0.02}^{+0.02}$  &$2.25_{-0.71}^{+0.36}$  &$0.78_{-0.14}^{+0.19}$  &$17264$\\
 \hline 
AT2019dsg & $43.50_{-0.05}^{+0.05}$  &$4.25_{-0.01}^{+0.01}$  &$1.62_{-0.04}^{+0.04}$  &$43.05_{-0.14}^{+0.17}$  &$4.26_{-0.01}^{+0.01}$  &$2.16_{-0.26}^{+0.25}$  &$1.24_{-0.11}^{+0.18}$  &$164922$\\
 \hline 
AT2019eve & $42.89_{-0.12}^{+0.04}$  &$3.98_{-0.03}^{+0.02}$  &$1.99_{-0.04}^{+0.03}$  &$42.39_{-0.12}^{+0.17}$  &$4.02_{-0.02}^{+0.05}$  &$2.87_{-0.24}^{+0.09}$  &$2.39_{-0.50}^{+0.45}$  &$95999$\\
 \hline 
AT2020opy & $43.94_{-0.02}^{+0.02}$  &$4.20_{-0.01}^{+0.01}$  &$1.69_{-0.03}^{+0.03}$  &$43.55_{-0.04}^{+0.05}$  &$4.21_{-0.01}^{+0.01}$  &$2.91_{-0.15}^{+0.12}$  &$3.68_{-0.68}^{+0.77}$  &$162141$\\
 \hline 
AT2020vwl & $42.16_{-0.14}^{+0.12}$  &$3.79_{-0.03}^{+0.03}$  &$1.74_{-0.03}^{+0.03}$  &$45.02_{-0.34}^{+0.22}$  &$4.42_{-0.02}^{+0.02}$  &$0.22_{-0.16}^{+0.26}$  &$1.35_{-0.03}^{+0.03}$  &$120564$\\
 \hline 
AT2020ysg & $44.59_{-0.04}^{+0.05}$  &$4.25_{-0.01}^{+0.01}$  &$2.09_{-0.02}^{+0.02}$  &$43.90_{-0.10}^{+0.12}$  &$4.26_{-0.01}^{+0.01}$  &$2.74_{-0.36}^{+0.20}$  &$1.19_{-0.30}^{+0.36}$  &$65506$\\
 \hline 
AT2020yue & $43.95_{-0.02}^{+0.02}$  &$3.97_{-0.01}^{+0.01}$  &$1.82_{-0.04}^{+0.03}$  &$43.59_{-0.06}^{+0.06}$  &$3.99_{-0.01}^{+0.01}$  &$3.05_{-0.08}^{+0.06}$  &$4.57_{-0.60}^{+0.32}$  &$27955$\\
 \hline 
AT2021ehb & $42.72_{-0.05}^{+0.05}$  &$4.33_{-0.02}^{+0.01}$  &$1.67_{-0.07}^{+0.06}$  &$42.57_{-0.10}^{+0.07}$  &$4.34_{-0.01}^{+0.01}$  &$2.10_{-0.16}^{+0.23}$  &$0.87_{-0.07}^{+0.10}$  &$109363$\\
 \hline 
AT2021lo & $43.80_{-0.12}^{+0.16}$  &$4.10_{-0.07}^{+0.11}$  &$1.79_{-0.08}^{+0.08}$  &$43.54_{-0.11}^{+0.11}$  &$4.42_{-0.13}^{+0.23}$  &$3.26_{-0.31}^{+0.18}$  &$3.18_{-1.13}^{+1.21}$  &$53225$\\
 \hline 
AT2021nwa & $43.15_{-0.01}^{+0.01}$  &$4.45_{-0.01}^{+0.01}$  &$1.92_{-0.03}^{+0.02}$  &$42.47_{-0.05}^{+0.05}$  &$4.52_{-0.03}^{+0.03}$  &$2.95_{-0.07}^{+0.04}$  &$2.25_{-0.27}^{+0.24}$  &$88268$\\
 \hline 
AT2021uqv & $43.19_{-0.05}^{+0.06}$  &$4.04_{-0.03}^{+0.03}$  &$1.45_{-0.05}^{+0.05}$  &$43.27_{-0.04}^{+0.04}$  &$4.27_{-0.02}^{+0.02}$  &$2.00_{-0.09}^{+0.09}$  &$0.83_{-0.05}^{+0.05}$  &$198795$\\
 \hline 
AT2021yzv & $44.65_{-0.02}^{+0.02}$  &$4.30_{-0.01}^{+0.01}$  &$2.10_{-0.03}^{+0.03}$  &$43.99_{-0.15}^{+0.12}$  &$4.31_{-0.01}^{+0.01}$  &$3.10_{-0.21}^{+0.13}$  &$3.90_{-1.13}^{+0.77}$  &$60969$\\
 \hline 
AT2022dsb & $43.15_{-0.06}^{+0.06}$  &$4.60_{-0.12}^{+0.18}$  &$1.28_{-0.04}^{+0.04}$  &$41.99_{-0.02}^{+0.03}$  &$5.89_{-0.13}^{+0.08}$  &$1.74_{-1.15}^{+0.97}$  &$0.01_{-0.01}^{+0.02}$  &$33485$\\
 \hline 
AT2022hvp & $44.98_{-0.03}^{+0.03}$  &$4.49_{-0.03}^{+0.03}$  &$1.41_{-0.02}^{+0.02}$  &$43.20_{-0.07}^{+0.20}$  &$4.50_{-0.03}^{+0.03}$  &$2.46_{-0.77}^{+0.40}$  &$0.82_{-0.29}^{+0.48}$  &$35389$\\
 \hline 
AT2023cvb & $43.95_{-0.08}^{+0.06}$  &$4.28_{-0.02}^{+0.02}$  &$1.73_{-0.06}^{+0.04}$  &$43.60_{-0.19}^{+0.17}$  &$4.30_{-0.01}^{+0.01}$  &$2.40_{-0.24}^{+0.32}$  &$2.09_{-0.32}^{+0.66}$  &$62541$\\
 \hline 
AT2019azh & $43.79_{-0.02}^{+0.02}$  &$4.30_{-0.01}^{+0.01}$  &$1.80_{-0.01}^{+0.01}$  &$41.87_{-0.02}^{+0.03}$  &$4.30_{-0.01}^{+0.01}$  &$1.76_{-1.18}^{+0.91}$  &$0.01_{-0.01}^{+0.02}$  &$212697$\\
 \hline 
AT2020mot & $43.56_{-0.01}^{+0.01}$  &$4.22_{-0.01}^{+0.01}$  &$1.73_{-0.02}^{+0.02}$  &$43.07_{-0.07}^{+0.07}$  &$4.23_{-0.01}^{+0.01}$  &$2.06_{-0.15}^{+0.14}$  &$0.95_{-0.07}^{+0.08}$  &$153718$\\
 \hline 
AT2020wey & $42.81_{-0.02}^{+0.02}$  &$4.29_{-0.02}^{+0.02}$  &$0.81_{-0.06}^{+0.05}$  &$42.09_{-0.08}^{+0.11}$  &$4.31_{-0.02}^{+0.02}$  &$2.10_{-0.28}^{+0.18}$  &$3.44_{-1.02}^{+1.05}$  &$99364$\\
 \hline 
AT2021axu & $43.74_{-0.03}^{+0.04}$  &$4.01_{-0.01}^{+0.01}$  &$1.78_{-0.02}^{+0.02}$  &$44.26_{-0.10}^{+0.08}$  &$5.83_{-0.20}^{+0.12}$  &$1.68_{-0.14}^{+0.19}$  &$1.20_{-0.08}^{+0.13}$  &$154420$\\
 \hline 
AT2021yte & $42.93_{-0.14}^{+0.12}$  &$4.07_{-0.06}^{+0.06}$  &$1.63_{-0.08}^{+0.08}$  &$42.91_{-0.39}^{+0.31}$  &$5.42_{-0.38}^{+0.38}$  &$0.73_{-0.48}^{+0.72}$  &$0.61_{-0.09}^{+0.10}$  &$53086$\\
 \hline 
ASASSN-14ae & $43.63_{-0.02}^{+0.02}$  &$4.21_{-0.02}^{+0.02}$  &$1.42_{-0.01}^{+0.01}$  &$42.53_{-0.35}^{+0.22}$  &$5.45_{-0.43}^{+0.37}$  &$0.70_{-0.42}^{+0.74}$  &$0.59_{-0.06}^{+0.08}$  &$9852$\\
 \hline 
iPTF-16fnl & $42.79_{-0.02}^{+0.02}$  &$4.32_{-0.02}^{+0.02}$  &$1.22_{-0.02}^{+0.02}$  &$41.42_{-0.03}^{+0.04}$  &$4.90_{-0.22}^{+0.45}$  &$1.77_{-1.20}^{+0.93}$  &$0.00_{-0.00}^{+0.01}$  &$17936$\\
 \hline 
AT2018hyz & $43.46_{-0.07}^{+0.06}$  &$4.04_{-0.02}^{+0.02}$  &$1.78_{-0.02}^{+0.02}$  &$43.80_{-0.04}^{+0.04}$  &$4.38_{-0.03}^{+0.04}$  &$1.25_{-0.09}^{+0.08}$  &$1.13_{-0.06}^{+0.06}$  &$6803$\\
 \hline 
AT2018lna & $43.70_{-0.03}^{+0.03}$  &$4.42_{-0.02}^{+0.02}$  &$1.72_{-0.03}^{+0.04}$  &$42.77_{-0.38}^{+0.40}$  &$4.48_{-0.04}^{+0.06}$  &$1.52_{-0.81}^{+0.92}$  &$0.67_{-0.15}^{+0.23}$  &$118645$\\
 \hline 
AT2018lni & $43.99_{-0.10}^{+0.12}$  &$4.40_{-0.13}^{+0.26}$  &$1.77_{-0.04}^{+0.04}$  &$42.89_{-0.08}^{+0.08}$  &$5.61_{-0.33}^{+0.28}$  &$3.74_{-0.75}^{+0.38}$  &$2.08_{-1.44}^{+1.87}$  &$112977$\\
 \hline 
AT2018zr & $43.47_{-0.04}^{+0.03}$  &$3.98_{-0.01}^{+0.01}$  &$1.66_{-0.04}^{+0.05}$  &$43.02_{-0.05}^{+0.06}$  &$4.35_{-0.03}^{+0.03}$  &$2.59_{-0.16}^{+0.18}$  &$1.81_{-0.31}^{+0.50}$  &$16757$\\
 \hline 
AT2019qiz & $43.15_{-0.02}^{+0.02}$  &$4.17_{-0.01}^{+0.01}$  &$1.48_{-0.01}^{+0.01}$  &$41.43_{-0.01}^{+0.02}$  &$4.33_{-0.01}^{+0.02}$  &$1.63_{-1.14}^{+1.05}$  &$0.00_{-0.00}^{+0.01}$  &$139619$\\
 \hline 
AT2020qhs & $44.57_{-0.17}^{+0.10}$  &$4.30_{-0.01}^{+0.01}$  &$1.93_{-0.03}^{+0.03}$  &$44.49_{-0.16}^{+0.15}$  &$4.31_{-0.01}^{+0.01}$  &$2.54_{-0.24}^{+0.28}$  &$2.03_{-0.36}^{+0.57}$  &$135089$\\
 \hline 
AT2020zso & $43.50_{-0.03}^{+0.02}$  &$4.23_{-0.02}^{+0.02}$  &$1.52_{-0.04}^{+0.04}$  &$41.97_{-0.14}^{+0.30}$  &$4.86_{-0.28}^{+0.58}$  &$2.22_{-1.06}^{+0.58}$  &$0.52_{-0.24}^{+0.51}$  &$58707$\\
 \hline 
ASASSN-15oi & $43.93_{-0.02}^{+0.02}$  &$4.54_{-0.04}^{+0.04}$  &$1.42_{-0.01}^{+0.01}$  &$41.83_{-0.31}^{+0.30}$  &$5.63_{-0.36}^{+0.26}$  &$1.12_{-0.77}^{+1.08}$  &$0.39_{-0.07}^{+0.09}$  &$22644$\\
 \hline 
AT2022upj & $43.17_{-0.05}^{+0.04}$  &$4.23_{-0.01}^{+0.01}$  &$1.99_{-0.19}^{+0.15}$  &$42.83_{-0.17}^{+0.14}$  &$4.24_{-0.01}^{+0.01}$  &$2.49_{-0.78}^{+0.38}$  &$0.58_{-0.34}^{+0.40}$  &$40718$\\
 \hline 
AT2021gje & $44.22_{-0.06}^{+0.06}$  &$4.13_{-0.03}^{+0.03}$  &$1.95_{-0.02}^{+0.02}$  &$44.06_{-0.26}^{+0.16}$  &$4.32_{-0.04}^{+0.03}$  &$0.44_{-0.31}^{+0.48}$  &$0.26_{-0.08}^{+0.06}$  &$190370$\\
 \hline 
AT2021jsg & $43.22_{-0.05}^{+0.05}$  &$4.04_{-0.02}^{+0.02}$  &$1.69_{-0.05}^{+0.04}$  &$42.84_{-0.21}^{+0.20}$  &$4.50_{-0.15}^{+0.30}$  &$2.44_{-1.15}^{+0.39}$  &$1.75_{-1.02}^{+2.06}$  &$52398$\\
 \hline 
AT2022lri & $41.76_{-0.11}^{+0.10}$  &$3.86_{-0.03}^{+0.03}$  &$2.15_{-0.04}^{+0.04}$  &$42.88_{-0.06}^{+0.06}$  &$5.62_{-0.32}^{+0.26}$  &$1.35_{-0.09}^{+0.09}$  &$1.14_{-0.04}^{+0.04}$  &$48536$\\
 \hline 
AT2023mhs & $42.09_{-0.92}^{+0.91}$  &$3.66_{-0.16}^{+0.20}$  &$1.57_{-0.12}^{+0.10}$  &$44.32_{-0.08}^{+0.08}$  &$4.22_{-0.02}^{+0.03}$  &$1.21_{-0.12}^{+0.11}$  &$1.98_{-0.15}^{+0.15}$  &$12923$\\
 \hline 
AT2022gri & $43.03_{-0.01}^{+0.02}$  &$4.44_{-0.01}^{+0.01}$  &$2.57_{-0.03}^{+0.03}$  &$42.58_{-0.13}^{+0.08}$  &$4.44_{-0.01}^{+0.01}$  &$2.61_{-0.67}^{+0.29}$  &$0.47_{-0.18}^{+0.28}$  &$28549$\\
 \hline

    \end{tabular}
    \caption{Same as \cref{tab:one power law}, but using the early-time exponential decay model  (Eq. \ref{eq: early model exp}) and the power-law plateau model (Eq. \ref{eq: late model PL}).}
    \label{tab:exp power-law}
\end{table}

\begin{table}[]
    \centering
    \begin{tabular}{|c|c|c|c|c|c|c|c|c|c|}
        
\hline 
TDE Name & $\log(\nu L_{\rm peak})$ &$\log(T_{\rm early})$ &$\log(t_{0,\rm decay})$ &$p_{\rm decay}$ &$\log(\nu L_{\rm plat})$ &$\log(T_{\rm plat})$ &AIC \\ 
 \hline 
 & $\log(\rm{erg}/\rm{s})$ &$\log(\rm{K})$ &$\log(s)$ &$1$ &$\log(\rm{erg}/\rm{s})$ &$\log(\rm{K})$ &\\ 
 \hline 
ASASSN-14li & $43.07_{-0.01}^{+0.02}$  &$4.55_{-0.02}^{+0.02}$  &$2.03_{-0.06}^{+0.06}$  &$3.14_{-0.24}^{+0.30}$  &$41.41_{-0.01}^{+0.01}$  &$5.90_{-0.12}^{+0.07}$  &$52471$\\
 \hline 
AT2018dyb & $43.66_{-0.02}^{+0.02}$  &$4.31_{-0.01}^{+0.01}$  &$1.46_{-0.04}^{+0.04}$  &$1.71_{-0.06}^{+0.06}$  &$37.33_{-1.61}^{+1.60}$  &$5.16_{-0.58}^{+0.56}$  &$7415$\\
 \hline 
AT2018hco & $44.04_{-0.05}^{+0.05}$  &$4.30_{-0.01}^{+0.01}$  &$1.56_{-0.11}^{+0.14}$  &$1.14_{-0.07}^{+0.13}$  &$41.41_{-1.87}^{+0.34}$  &$4.52_{-0.16}^{+0.77}$  &$179456$\\
 \hline 
AT2018jbv & $44.88_{-0.04}^{+0.04}$  &$4.31_{-0.01}^{+0.01}$  &$2.39_{-0.11}^{+0.11}$  &$2.36_{-0.37}^{+0.44}$  &$43.35_{-0.08}^{+0.06}$  &$4.32_{-0.01}^{+0.01}$  &$124468$\\
 \hline 
AT2019ahk & $43.86_{-0.01}^{+0.01}$  &$4.20_{-0.01}^{+0.01}$  &$1.84_{-0.05}^{+0.05}$  &$2.02_{-0.12}^{+0.13}$  &$41.75_{-0.03}^{+0.04}$  &$5.36_{-0.37}^{+0.42}$  &$17253$\\
 \hline 
AT2019dsg & $43.53_{-0.03}^{+0.04}$  &$4.25_{-0.01}^{+0.01}$  &$1.51_{-0.05}^{+0.05}$  &$1.08_{-0.04}^{+0.04}$  &$37.60_{-1.80}^{+2.05}$  &$5.08_{-0.65}^{+0.63}$  &$164894$\\
 \hline 
AT2019eve & $43.00_{-0.03}^{+0.03}$  &$3.98_{-0.01}^{+0.01}$  &$2.22_{-0.07}^{+0.08}$  &$2.09_{-0.20}^{+0.24}$  &$41.30_{-0.32}^{+0.16}$  &$5.48_{-0.50}^{+0.36}$  &$95999$\\
 \hline 
AT2020opy & $44.04_{-0.01}^{+0.01}$  &$4.20_{-0.01}^{+0.01}$  &$1.98_{-0.04}^{+0.04}$  &$1.64_{-0.07}^{+0.07}$  &$38.33_{-2.28}^{+2.18}$  &$5.11_{-0.61}^{+0.60}$  &$162118$\\
 \hline 
AT2020vwl & $43.69_{-0.06}^{+0.07}$  &$4.26_{-0.01}^{+0.01}$  &$1.46_{-0.08}^{+0.08}$  &$1.76_{-0.08}^{+0.08}$  &$41.03_{-0.05}^{+0.05}$  &$5.76_{-0.26}^{+0.17}$  &$120533$\\
 \hline 
AT2020ysg & $44.63_{-0.03}^{+0.05}$  &$4.25_{-0.01}^{+0.01}$  &$2.72_{-0.09}^{+0.05}$  &$4.39_{-0.69}^{+0.44}$  &$43.23_{-0.04}^{+0.03}$  &$4.27_{-0.02}^{+0.02}$  &$65513$\\
 \hline 
AT2020yue & $44.07_{-0.01}^{+0.01}$  &$3.98_{-0.00}^{+0.00}$  &$2.16_{-0.04}^{+0.04}$  &$1.94_{-0.10}^{+0.12}$  &$38.39_{-2.34}^{+2.30}$  &$4.99_{-0.69}^{+0.68}$  &$27956$\\
 \hline 
AT2021ehb & $42.93_{-0.03}^{+0.04}$  &$4.34_{-0.01}^{+0.01}$  &$1.43_{-0.08}^{+0.07}$  &$0.75_{-0.03}^{+0.03}$  &$37.61_{-1.77}^{+1.80}$  &$5.10_{-0.60}^{+0.62}$  &$109357$\\
 \hline 
AT2021lo & $44.10_{-0.10}^{+0.10}$  &$4.27_{-0.08}^{+0.11}$  &$1.62_{-0.15}^{+0.16}$  &$0.93_{-0.08}^{+0.10}$  &$36.54_{-4.55}^{+4.63}$  &$5.17_{-0.60}^{+0.57}$  &$53219$\\
 \hline 
AT2021nwa & $43.20_{-0.01}^{+0.01}$  &$4.46_{-0.01}^{+0.01}$  &$2.36_{-0.09}^{+0.09}$  &$2.80_{-0.37}^{+0.48}$  &$41.58_{-0.07}^{+0.06}$  &$4.49_{-0.02}^{+0.03}$  &$88292$\\
 \hline 
AT2021uqv & $43.48_{-0.03}^{+0.02}$  &$4.14_{-0.02}^{+0.01}$  &$1.45_{-0.05}^{+0.05}$  &$0.88_{-0.04}^{+0.04}$  &$42.24_{-0.08}^{+0.07}$  &$5.73_{-0.31}^{+0.20}$  &$198768$\\
 \hline 
AT2021yzv & $44.72_{-0.01}^{+0.01}$  &$4.31_{-0.01}^{+0.01}$  &$2.61_{-0.07}^{+0.08}$  &$3.38_{-0.35}^{+0.45}$  &$38.59_{-2.43}^{+2.59}$  &$5.13_{-0.57}^{+0.58}$  &$60968$\\
 \hline 
AT2022dsb & $43.26_{-0.08}^{+0.09}$  &$4.61_{-0.12}^{+0.18}$  &$1.74_{-0.11}^{+0.07}$  &$4.48_{-0.65}^{+0.37}$  &$41.97_{-0.02}^{+0.02}$  &$5.90_{-0.13}^{+0.07}$  &$33483$\\
 \hline 
AT2022hvp & $45.03_{-0.04}^{+0.04}$  &$4.53_{-0.03}^{+0.03}$  &$1.94_{-0.07}^{+0.05}$  &$4.48_{-0.42}^{+0.35}$  &$42.79_{-0.03}^{+0.02}$  &$4.54_{-0.03}^{+0.04}$  &$35405$\\
 \hline 
AT2023cvb & $44.23_{-0.05}^{+0.06}$  &$4.30_{-0.01}^{+0.01}$  &$1.77_{-0.09}^{+0.08}$  &$1.67_{-0.08}^{+0.09}$  &$39.30_{-2.94}^{+2.02}$  &$5.14_{-0.63}^{+0.61}$  &$62537$\\
 \hline 
AT2019azh & $43.87_{-0.02}^{+0.02}$  &$4.30_{-0.01}^{+0.01}$  &$2.27_{-0.02}^{+0.01}$  &$4.95_{-0.09}^{+0.04}$  &$41.83_{-0.01}^{+0.01}$  &$4.30_{-0.01}^{+0.01}$  &$212708$\\
 \hline 
AT2020mot & $43.62_{-0.01}^{+0.01}$  &$4.23_{-0.01}^{+0.01}$  &$1.91_{-0.06}^{+0.06}$  &$1.78_{-0.12}^{+0.14}$  &$41.98_{-0.04}^{+0.04}$  &$4.25_{-0.02}^{+0.03}$  &$153682$\\
 \hline 
AT2020wey & $42.83_{-0.02}^{+0.02}$  &$4.29_{-0.01}^{+0.01}$  &$0.89_{-0.07}^{+0.07}$  &$1.47_{-0.09}^{+0.11}$  &$39.72_{-3.03}^{+0.54}$  &$5.30_{-0.57}^{+0.47}$  &$99358$\\
 \hline 
AT2021axu & $44.37_{-0.01}^{+0.01}$  &$4.39_{-0.01}^{+0.01}$  &$2.04_{-0.05}^{+0.06}$  &$2.41_{-0.20}^{+0.23}$  &$42.51_{-0.10}^{+0.08}$  &$5.64_{-0.36}^{+0.25}$  &$154407$\\
 \hline 
AT2021yte & $43.27_{-0.05}^{+0.05}$  &$4.27_{-0.02}^{+0.03}$  &$1.88_{-0.17}^{+0.19}$  &$2.70_{-0.52}^{+0.83}$  &$41.59_{-0.07}^{+0.06}$  &$5.56_{-0.37}^{+0.30}$  &$53080$\\
 \hline 
ASASSN-14ae & $43.68_{-0.02}^{+0.02}$  &$4.25_{-0.01}^{+0.01}$  &$1.91_{-0.07}^{+0.07}$  &$4.07_{-0.40}^{+0.47}$  &$41.14_{-0.04}^{+0.04}$  &$5.49_{-0.45}^{+0.34}$  &$9858$\\
 \hline 
iPTF-16fnl & $42.88_{-0.03}^{+0.03}$  &$4.32_{-0.02}^{+0.02}$  &$1.72_{-0.06}^{+0.04}$  &$4.66_{-0.42}^{+0.25}$  &$41.38_{-0.02}^{+0.03}$  &$5.28_{-0.38}^{+0.48}$  &$17933$\\
 \hline 
AT2018hyz & $43.96_{-0.01}^{+0.01}$  &$4.21_{-0.00}^{+0.00}$  &$1.72_{-0.05}^{+0.05}$  &$1.81_{-0.09}^{+0.12}$  &$41.30_{-0.07}^{+0.06}$  &$5.52_{-0.41}^{+0.35}$  &$6803$\\
 \hline 
AT2018lna & $43.74_{-0.02}^{+0.02}$  &$4.44_{-0.02}^{+0.02}$  &$2.22_{-0.11}^{+0.10}$  &$3.89_{-0.62}^{+0.66}$  &$41.77_{-0.05}^{+0.05}$  &$4.49_{-0.04}^{+0.07}$  &$118644$\\
 \hline 
AT2018lni & $44.07_{-0.10}^{+0.12}$  &$4.48_{-0.14}^{+0.35}$  &$2.24_{-0.17}^{+0.10}$  &$4.03_{-0.92}^{+0.70}$  &$42.67_{-0.03}^{+0.03}$  &$5.66_{-0.36}^{+0.24}$  &$112984$\\
 \hline 
AT2018zr & $43.63_{-0.03}^{+0.03}$  &$4.10_{-0.01}^{+0.01}$  &$1.86_{-0.13}^{+0.13}$  &$1.44_{-0.16}^{+0.22}$  &$41.70_{-0.09}^{+0.07}$  &$5.54_{-0.42}^{+0.32}$  &$16763$\\
 \hline 
AT2019qiz & $43.22_{-0.02}^{+0.02}$  &$4.17_{-0.01}^{+0.01}$  &$1.95_{-0.05}^{+0.03}$  &$4.68_{-0.35}^{+0.23}$  &$41.42_{-0.01}^{+0.01}$  &$4.35_{-0.02}^{+0.02}$  &$139625$\\
 \hline 
AT2020qhs & $44.85_{-0.01}^{+0.01}$  &$4.30_{-0.01}^{+0.01}$  &$2.26_{-0.04}^{+0.06}$  &$2.00_{-0.10}^{+0.19}$  &$39.78_{-3.27}^{+2.76}$  &$4.92_{-0.55}^{+0.71}$  &$135090$\\
 \hline 
AT2020zso & $43.53_{-0.02}^{+0.02}$  &$4.24_{-0.02}^{+0.02}$  &$2.03_{-0.12}^{+0.08}$  &$4.18_{-0.68}^{+0.56}$  &$41.51_{-0.09}^{+0.07}$  &$5.11_{-0.44}^{+0.55}$  &$58707$\\
 \hline 
ASASSN-15oi & $44.19_{-0.04}^{+0.04}$  &$4.58_{-0.03}^{+0.04}$  &$1.68_{-0.07}^{+0.07}$  &$3.63_{-0.24}^{+0.29}$  &$41.04_{-0.04}^{+0.04}$  &$5.61_{-0.35}^{+0.28}$  &$22649$\\
 \hline 
AT2022upj & $43.25_{-0.03}^{+0.03}$  &$4.23_{-0.01}^{+0.01}$  &$2.22_{-0.38}^{+0.36}$  &$1.76_{-0.86}^{+1.55}$  &$42.33_{-0.33}^{+0.10}$  &$4.25_{-0.01}^{+0.03}$  &$40715$\\
 \hline 
AT2021gje & $44.38_{-0.03}^{+0.03}$  &$4.17_{-0.02}^{+0.02}$  &$2.15_{-0.27}^{+0.22}$  &$2.36_{-0.78}^{+1.08}$  &$43.36_{-0.06}^{+0.05}$  &$4.35_{-0.06}^{+0.07}$  &$190393$\\
 \hline 
AT2021jsg & $43.36_{-0.03}^{+0.03}$  &$4.11_{-0.02}^{+0.02}$  &$2.20_{-0.11}^{+0.10}$  &$3.72_{-0.69}^{+0.77}$  &$41.99_{-0.11}^{+0.09}$  &$5.59_{-0.37}^{+0.29}$  &$52392$\\
 \hline 
AT2022lri & $42.77_{-0.02}^{+0.03}$  &$4.31_{-0.01}^{+0.01}$  &$2.48_{-0.11}^{+0.09}$  &$4.02_{-0.62}^{+0.61}$  &$41.12_{-0.03}^{+0.02}$  &$5.74_{-0.27}^{+0.18}$  &$48500$\\
 \hline 
AT2023mhs & $44.11_{-0.05}^{+0.04}$  &$4.12_{-0.01}^{+0.01}$  &$1.71_{-0.08}^{+0.08}$  &$3.00_{-0.28}^{+0.33}$  &$41.13_{-0.36}^{+0.15}$  &$5.30_{-0.53}^{+0.49}$  &$12914$\\
 \hline 
AT2022gri & $43.11_{-0.01}^{+0.01}$  &$4.44_{-0.01}^{+0.01}$  &$2.92_{-0.06}^{+0.05}$  &$2.09_{-0.21}^{+0.20}$  &$37.59_{-1.75}^{+1.79}$  &$5.18_{-0.52}^{+0.57}$  &$28500$\\
 \hline

    \end{tabular}
    \caption{Same as \cref{tab:one power law}, but using the early-time power-law decay model  (Eq. \ref{eq: early model PL}) and the flat plateau model (Eq. \ref{eq: late model flat}).}
    \label{tab:power-law flat}
\end{table}

\begin{table}[]
    \centering
    \begin{tabular}{|c|c|c|c|c|c|c|c|c|}
   
       \hline 
TDE Name & $\log(\nu L_{\rm peak})$ &$\log(T_{\rm early})$ &$\log(t_{0,\rm decay})$ &$p_{\rm decay}$ &$\log(\nu L_{plat})$ &$\log(T_{\rm plat})$ &$\log(\tau_{\rm cuesta})$ &AIC \\ 
 \hline 
 & $\log(\rm{erg}/\rm{s})$ &$\log(\rm{K})$ &$\log(s)$ &$1$ &$\log(\rm{erg}/\rm{s})$ &$\log(\rm{K})$ &$s$ &\\ 
 \hline 
ASASSN-14li & $43.07_{-0.01}^{+0.01}$  &$4.53_{-0.02}^{+0.02}$  &$2.26_{-0.01}^{+0.01}$  &$4.95_{-0.09}^{+0.04}$  &$41.58_{-0.01}^{+0.01}$  &$5.92_{-0.11}^{+0.06}$  &$3.48_{-0.03}^{+0.03}$  &$52298$\\
 \hline 
AT2018dyb & $43.71_{-0.02}^{+0.02}$  &$4.30_{-0.01}^{+0.01}$  &$2.12_{-0.03}^{+0.02}$  &$4.88_{-0.20}^{+0.09}$  &$42.09_{-0.03}^{+0.03}$  &$4.33_{-0.01}^{+0.02}$  &$2.67_{-0.04}^{+0.03}$  &$7376$\\
 \hline 
AT2018hco & $43.89_{-0.04}^{+0.04}$  &$4.28_{-0.01}^{+0.01}$  &$2.21_{-0.20}^{+0.22}$  &$2.66_{-0.72}^{+1.15}$  &$42.67_{-0.15}^{+0.09}$  &$4.33_{-0.03}^{+0.04}$  &$3.18_{-0.06}^{+0.10}$  &$179471$\\
 \hline 
AT2018jbv & $44.88_{-0.03}^{+0.03}$  &$4.31_{-0.01}^{+0.01}$  &$2.64_{-0.15}^{+0.08}$  &$4.06_{-1.12}^{+0.68}$  &$43.71_{-0.18}^{+0.09}$  &$4.32_{-0.01}^{+0.01}$  &$3.41_{-0.13}^{+0.35}$  &$124469$\\
 \hline 
AT2019ahk & $43.85_{-0.01}^{+0.01}$  &$4.20_{-0.01}^{+0.01}$  &$1.98_{-0.09}^{+0.10}$  &$2.48_{-0.31}^{+0.39}$  &$41.99_{-0.13}^{+0.10}$  &$4.70_{-0.13}^{+0.32}$  &$3.60_{-0.14}^{+0.25}$  &$17248$\\
 \hline 
AT2019dsg & $44.72_{-0.74}^{+0.84}$  &$4.25_{-0.01}^{+0.01}$  &$0.91_{-0.63}^{+0.48}$  &$1.38_{-0.13}^{+0.07}$  &$41.33_{-0.25}^{+0.12}$  &$4.27_{-0.02}^{+0.05}$  &$6.41_{-2.09}^{+2.38}$  &$164881$\\
 \hline 
AT2019eve & $42.92_{-0.04}^{+0.04}$  &$3.94_{-0.02}^{+0.02}$  &$2.14_{-0.08}^{+0.08}$  &$1.91_{-0.17}^{+0.21}$  &$41.92_{-0.14}^{+0.12}$  &$5.34_{-0.52}^{+0.46}$  &$2.79_{-0.16}^{+0.18}$  &$95991$\\
 \hline 
AT2020opy & $43.56_{-0.05}^{+0.08}$  &$4.00_{-0.02}^{+0.02}$  &$2.19_{-0.08}^{+0.07}$  &$1.73_{-0.11}^{+0.12}$  &$43.89_{-0.07}^{+0.03}$  &$4.58_{-0.05}^{+0.12}$  &$1.76_{-0.03}^{+0.03}$  &$162087$\\
 \hline 
AT2020vwl & $43.67_{-0.06}^{+0.07}$  &$4.21_{-0.01}^{+0.01}$  &$1.46_{-0.08}^{+0.08}$  &$1.87_{-0.09}^{+0.10}$  &$41.67_{-0.08}^{+0.08}$  &$5.74_{-0.28}^{+0.18}$  &$2.87_{-0.06}^{+0.07}$  &$120517$\\
 \hline 
AT2020ysg & $44.63_{-0.03}^{+0.06}$  &$4.25_{-0.01}^{+0.01}$  &$2.72_{-0.07}^{+0.04}$  &$4.50_{-0.60}^{+0.36}$  &$43.27_{-0.06}^{+0.13}$  &$4.27_{-0.01}^{+0.02}$  &$4.57_{-1.12}^{+3.63}$  &$65514$\\
 \hline 
AT2020yue & $44.03_{-0.02}^{+0.02}$  &$3.97_{-0.01}^{+0.01}$  &$2.30_{-0.13}^{+0.15}$  &$3.00_{-0.81}^{+1.22}$  &$43.20_{-0.26}^{+0.15}$  &$4.00_{-0.02}^{+0.03}$  &$2.60_{-0.07}^{+0.08}$  &$27955$\\
 \hline 
AT2021ehb & $42.92_{-0.03}^{+0.04}$  &$4.33_{-0.06}^{+0.01}$  &$1.37_{-0.36}^{+0.11}$  &$0.73_{-0.11}^{+0.04}$  &$38.42_{-5.69}^{+3.50}$  &$5.36_{-0.71}^{+0.45}$  &$3.97_{-1.54}^{+4.10}$  &$109365$\\
 \hline 
AT2021lo & $43.72_{-0.17}^{+0.17}$  &$4.02_{-0.07}^{+0.10}$  &$2.15_{-0.27}^{+0.24}$  &$2.76_{-1.03}^{+1.54}$  &$43.42_{-0.13}^{+0.10}$  &$4.86_{-0.39}^{+0.71}$  &$3.02_{-0.07}^{+0.10}$  &$53230$\\
 \hline 
AT2021nwa & $43.20_{-0.02}^{+0.02}$  &$4.47_{-0.01}^{+0.01}$  &$1.92_{-0.07}^{+0.07}$  &$1.39_{-0.10}^{+0.11}$  &$42.34_{-0.65}^{+0.18}$  &$4.52_{-0.04}^{+0.38}$  &$1.90_{-0.09}^{+0.11}$  &$88271$\\
 \hline 
AT2021uqv & $43.20_{-0.03}^{+0.03}$  &$3.98_{-0.01}^{+0.01}$  &$0.88_{-0.06}^{+0.08}$  &$0.56_{-0.02}^{+0.02}$  &$43.12_{-0.03}^{+0.04}$  &$5.77_{-0.25}^{+0.16}$  &$2.05_{-0.03}^{+0.03}$  &$198718$\\
 \hline 
AT2021yzv & $44.71_{-0.32}^{+0.02}$  &$4.30_{-0.19}^{+0.01}$  &$2.63_{-0.08}^{+0.09}$  &$3.40_{-0.38}^{+0.56}$  &$39.56_{-6.75}^{+4.91}$  &$5.31_{-0.75}^{+0.46}$  &$3.50_{-1.48}^{+4.35}$  &$60978$\\
 \hline 
AT2022dsb & $43.26_{-0.08}^{+0.09}$  &$4.62_{-0.12}^{+0.18}$  &$1.74_{-0.12}^{+0.08}$  &$4.46_{-0.68}^{+0.39}$  &$41.97_{-0.02}^{+0.02}$  &$5.90_{-0.13}^{+0.07}$  &$7.67_{-1.80}^{+1.65}$  &$33485$\\
 \hline 
AT2022hvp & $45.05_{-0.04}^{+0.04}$  &$4.52_{-0.03}^{+0.03}$  &$1.95_{-0.04}^{+0.03}$  &$4.78_{-0.32}^{+0.17}$  &$42.91_{-0.08}^{+0.05}$  &$4.53_{-0.03}^{+0.04}$  &$3.35_{-0.15}^{+0.46}$  &$35397$\\
 \hline 
AT2023cvb & $44.12_{-0.04}^{+0.04}$  &$4.29_{-0.02}^{+0.01}$  &$2.30_{-0.12}^{+0.10}$  &$3.90_{-0.75}^{+0.76}$  &$42.85_{-0.14}^{+0.10}$  &$4.31_{-0.01}^{+0.02}$  &$2.77_{-0.07}^{+0.09}$  &$62544$\\
 \hline 
AT2019azh & $43.88_{-0.02}^{+0.02}$  &$4.30_{-0.01}^{+0.01}$  &$2.27_{-0.01}^{+0.01}$  &$4.95_{-0.09}^{+0.04}$  &$41.83_{-0.01}^{+0.01}$  &$4.30_{-0.01}^{+0.01}$  &$7.72_{-1.57}^{+1.51}$  &$212710$\\
 \hline 
AT2020mot & $43.63_{-0.01}^{+0.01}$  &$4.23_{-0.01}^{+0.01}$  &$1.91_{-0.05}^{+0.06}$  &$1.79_{-0.12}^{+0.14}$  &$41.99_{-0.04}^{+0.04}$  &$4.26_{-0.02}^{+0.03}$  &$7.20_{-2.00}^{+1.94}$  &$153684$\\
 \hline 
AT2020wey & $42.83_{-0.02}^{+0.02}$  &$4.29_{-0.01}^{+0.02}$  &$0.87_{-0.07}^{+0.07}$  &$1.44_{-0.09}^{+0.10}$  &$40.21_{-2.99}^{+8.23}$  &$5.21_{-0.59}^{+0.54}$  &$-2.17_{-5.49}^{+9.12}$  &$99363$\\
 \hline 
AT2021axu & $43.94_{-0.02}^{+0.02}$  &$4.06_{-0.01}^{+0.01}$  &$2.31_{-0.02}^{+0.01}$  &$4.87_{-0.20}^{+0.10}$  &$43.81_{-0.02}^{+0.02}$  &$5.89_{-0.14}^{+0.08}$  &$2.39_{-0.02}^{+0.02}$  &$154464$\\
 \hline 
AT2021yte & $43.26_{-0.05}^{+0.05}$  &$4.27_{-0.03}^{+0.02}$  &$1.90_{-0.17}^{+0.19}$  &$2.77_{-0.56}^{+0.90}$  &$41.61_{-0.07}^{+0.07}$  &$5.59_{-0.37}^{+0.28}$  &$6.44_{-2.31}^{+2.38}$  &$53082$\\
 \hline 
ASASSN-14ae & $43.68_{-0.02}^{+0.02}$  &$4.24_{-0.01}^{+0.01}$  &$1.98_{-0.06}^{+0.03}$  &$4.62_{-0.41}^{+0.27}$  &$41.34_{-0.06}^{+0.05}$  &$5.51_{-0.45}^{+0.34}$  &$3.56_{-0.11}^{+0.15}$  &$9847$\\
 \hline 
iPTF-16fnl & $42.87_{-0.03}^{+0.03}$  &$4.32_{-0.02}^{+0.02}$  &$1.72_{-0.06}^{+0.04}$  &$4.65_{-0.41}^{+0.26}$  &$41.38_{-0.02}^{+0.03}$  &$5.26_{-0.35}^{+0.48}$  &$7.81_{-1.49}^{+1.50}$  &$17935$\\
 \hline 
AT2018hyz & $43.95_{-0.01}^{+0.01}$  &$4.19_{-0.01}^{+0.01}$  &$1.83_{-0.06}^{+0.06}$  &$2.18_{-0.17}^{+0.20}$  &$41.94_{-0.11}^{+0.10}$  &$5.58_{-0.39}^{+0.29}$  &$3.01_{-0.07}^{+0.09}$  &$6797$\\
 \hline 
AT2018lna & $43.74_{-0.02}^{+0.02}$  &$4.44_{-0.02}^{+0.02}$  &$2.23_{-0.11}^{+0.09}$  &$3.99_{-0.66}^{+0.64}$  &$41.79_{-0.06}^{+0.09}$  &$4.49_{-0.04}^{+0.07}$  &$5.71_{-1.98}^{+2.75}$  &$118646$\\
 \hline 
AT2018lni & $44.07_{-0.10}^{+0.10}$  &$4.46_{-0.14}^{+0.26}$  &$2.24_{-0.14}^{+0.09}$  &$4.18_{-0.83}^{+0.58}$  &$42.72_{-0.06}^{+0.09}$  &$5.64_{-0.37}^{+0.26}$  &$4.35_{-0.72}^{+3.37}$  &$112984$\\
 \hline 
AT2018zr & $43.63_{-0.03}^{+0.03}$  &$4.03_{-0.01}^{+0.01}$  &$1.58_{-0.18}^{+0.17}$  &$1.34_{-0.21}^{+0.28}$  &$42.50_{-0.05}^{+0.05}$  &$5.45_{-0.45}^{+0.38}$  &$2.66_{-0.05}^{+0.05}$  &$16772$\\
 \hline 
AT2019qiz & $43.22_{-0.02}^{+0.02}$  &$4.17_{-0.01}^{+0.01}$  &$1.95_{-0.05}^{+0.03}$  &$4.69_{-0.33}^{+0.22}$  &$41.42_{-0.01}^{+0.01}$  &$4.34_{-0.02}^{+0.02}$  &$7.84_{-1.51}^{+1.47}$  &$139627$\\
 \hline 
AT2020qhs & $44.81_{-0.01}^{+0.01}$  &$4.30_{-0.01}^{+0.01}$  &$2.54_{-0.09}^{+0.10}$  &$3.62_{-0.70}^{+0.92}$  &$43.58_{-0.26}^{+0.17}$  &$4.32_{-0.01}^{+0.02}$  &$3.04_{-0.13}^{+0.26}$  &$135088$\\
 \hline 
AT2020zso & $43.48_{-0.03}^{+0.03}$  &$4.28_{-0.02}^{+0.02}$  &$0.51_{-0.30}^{+0.36}$  &$0.94_{-0.13}^{+0.21}$  &$43.29_{-0.09}^{+0.07}$  &$4.28_{-0.02}^{+0.02}$  &$1.59_{-0.05}^{+0.05}$  &$58711$\\
 \hline 
ASASSN-15oi & $44.19_{-0.04}^{+0.04}$  &$4.58_{-0.03}^{+0.04}$  &$1.69_{-0.07}^{+0.08}$  &$3.68_{-0.28}^{+0.36}$  &$41.07_{-0.05}^{+0.06}$  &$5.65_{-0.37}^{+0.25}$  &$5.83_{-1.85}^{+2.80}$  &$22651$\\
 \hline 
AT2022upj & $43.26_{-0.03}^{+0.04}$  &$4.24_{-0.01}^{+0.01}$  &$2.15_{-0.32}^{+0.40}$  &$1.54_{-0.66}^{+1.53}$  &$42.32_{-0.30}^{+0.11}$  &$4.25_{-0.02}^{+0.03}$  &$6.48_{-2.53}^{+2.37}$  &$40718$\\
 \hline 
AT2021gje & $44.38_{-0.03}^{+0.03}$  &$4.17_{-0.02}^{+0.02}$  &$2.16_{-0.24}^{+0.22}$  &$2.38_{-0.71}^{+1.09}$  &$43.37_{-0.05}^{+0.04}$  &$4.35_{-0.05}^{+0.06}$  &$7.25_{-2.08}^{+1.88}$  &$190397$\\
 \hline 
AT2021jsg & $43.30_{-0.05}^{+0.04}$  &$4.07_{-0.03}^{+0.03}$  &$2.11_{-0.18}^{+0.15}$  &$3.14_{-0.83}^{+1.11}$  &$42.38_{-0.24}^{+0.31}$  &$5.49_{-0.47}^{+0.36}$  &$2.51_{-0.59}^{+0.73}$  &$52395$\\
 \hline 
AT2022lri & $42.77_{-0.02}^{+0.03}$  &$4.31_{-0.01}^{+0.01}$  &$2.48_{-0.11}^{+0.08}$  &$4.10_{-0.64}^{+0.60}$  &$41.14_{-0.03}^{+0.07}$  &$5.74_{-0.28}^{+0.18}$  &$5.81_{-2.17}^{+2.80}$  &$48503$\\
 \hline 
AT2023mhs & $44.62_{-0.29}^{+0.27}$  &$4.11_{-0.04}^{+0.02}$  &$0.41_{-0.27}^{+0.48}$  &$1.49_{-0.13}^{+1.09}$  &$43.53_{-0.05}^{+0.05}$  &$4.16_{-0.02}^{+0.02}$  &$1.72_{-0.05}^{+0.10}$  &$12935$\\
 \hline 
AT2022gri & $43.11_{-0.01}^{+0.01}$  &$4.44_{-0.01}^{+0.01}$  &$2.92_{-0.06}^{+0.05}$  &$2.09_{-0.20}^{+0.20}$  &$38.53_{-2.40}^{+2.84}$  &$5.20_{-0.52}^{+0.54}$  &$-1.11_{-5.93}^{+6.91}$  &$28502$\\
 \hline

    \end{tabular}
    \caption{Same as \cref{tab:one power law}, but using the early-time power-law decay model  (Eq. \ref{eq: early model PL}) and the exponential plateau model (Eq. \ref{eq: late model exp}).}
    \label{tab:power-law exp}
\end{table}

\section{Magnetically elevated disk fits }
\label{app:mag fits}
In \cref{tab:mag exp,tab:mag PL}, we show results for the 25 TDEs that we fit to a magnetically elevated disk model as in in \S\ref{subsec: mag fitting}. The early-time model (Eq. \ref{eq: early model exp} or Eq. \ref{eq: early model PL}) was chosen based on results from \S\ref{sec:pheno}.  

\begin{table}[]
    \centering
    \begin{tabular}{|c|c|c|c|c|c|c|c|}
\hline 
TDE Name & $\log(\nu L_{\rm peak})$ & $\log(T_{\rm early})$ & $\log(\tau_{\rm decay})$ & $\log(\alpha)$ & $\log(M)$ & $\log(m_\star)$ & $\cos(i)$ \\ 
 \hline 
 & $\log(\rm{erg}/\rm{s})$ & $\log(\rm{K})$ & $\log(s)$ & $1$ & $\log(M_\odot)$ & $\log(M_\odot)$ & $1$ \\ 
 \hline 
ASASSN-14li & $43.02_{-0.01}^{+0.01}$  &$4.54_{-0.02}^{+0.02}$  &$1.70_{-0.01}^{+0.01}$  &$-1.24_{-0.12}^{+0.14}$  &$6.78_{-0.24}^{+0.25}$  &$-0.52_{-0.17}^{+0.15}$  &$0.56_{-0.18}^{+0.25}$  \\
 \hline 
AT2018dyb & $43.72_{-0.04}^{+0.04}$  &$4.28_{-0.01}^{+0.02}$  &$1.65_{-0.01}^{+0.01}$  &$0.07_{-0.21}^{+0.19}$  &$6.45_{-0.64}^{+0.43}$  &$-0.36_{-0.21}^{+0.37}$  &$0.57_{-0.22}^{+0.26}$  \\
 \hline 
AT2021nwa & $43.18_{-0.01}^{+0.01}$  &$4.46_{-0.01}^{+0.01}$  &$2.04_{-0.01}^{+0.01}$  &$-0.48_{-0.22}^{+0.26}$  &$6.59_{-0.52}^{+0.38}$  &$-0.35_{-0.20}^{+0.34}$  &$0.65_{-0.31}^{+0.25}$  \\
 \hline 
AT2019azh & $43.80_{-0.02}^{+0.03}$  &$4.29_{-0.02}^{+0.02}$  &$1.79_{-0.01}^{+0.01}$  &$-0.84_{-0.25}^{+0.23}$  &$5.81_{-0.65}^{+0.51}$  &$0.11_{-0.32}^{+0.45}$  &$0.79_{-0.22}^{+0.16}$  \\
 \hline 
AT2021yte & $43.14_{-0.04}^{+0.04}$  &$4.24_{-0.02}^{+0.02}$  &$1.78_{-0.04}^{+0.03}$  &$-3.07_{-0.62}^{+0.62}$  &$6.80_{-0.61}^{+0.37}$  &$-0.08_{-0.27}^{+0.41}$  &$0.73_{-0.30}^{+0.19}$  \\
 \hline 
AT2018lna & $43.72_{-0.02}^{+0.02}$  &$4.41_{-0.02}^{+0.02}$  &$1.74_{-0.03}^{+0.04}$  &$-0.09_{-0.32}^{+0.25}$  &$6.42_{-0.62}^{+0.48}$  &$-0.21_{-0.24}^{+0.38}$  &$0.69_{-0.28}^{+0.21}$  \\
 \hline 
AT2019qiz & $43.14_{-0.02}^{+0.02}$  &$4.16_{-0.01}^{+0.01}$  &$1.50_{-0.01}^{+0.01}$  &$-2.01_{-0.24}^{+0.23}$  &$5.11_{-0.60}^{+0.59}$  &$0.07_{-0.37}^{+0.46}$  &$0.78_{-0.22}^{+0.16}$  \\
 \hline 
AT2020zso & $43.48_{-0.02}^{+0.02}$  &$4.25_{-0.02}^{+0.02}$  &$1.58_{-0.04}^{+0.03}$  &$-0.48_{-0.39}^{+0.33}$  &$6.03_{-0.77}^{+0.59}$  &$-0.43_{-0.35}^{+0.42}$  &$0.65_{-0.29}^{+0.24}$  \\
 \hline 
ASASSN-15oi & $43.92_{-0.02}^{+0.02}$  &$4.55_{-0.04}^{+0.05}$  &$1.43_{-0.01}^{+0.01}$  &$-1.53_{-0.26}^{+0.27}$  &$6.11_{-0.78}^{+0.53}$  &$-0.59_{-0.31}^{+0.40}$  &$0.68_{-0.29}^{+0.23}$  \\
 \hline 
AT2022lri & $42.68_{-0.02}^{+0.02}$  &$4.29_{-0.01}^{+0.01}$  &$2.05_{-0.04}^{+0.03}$  &$-2.15_{-0.94}^{+0.66}$  &$6.55_{-0.53}^{+0.35}$  &$-0.62_{-0.21}^{+0.31}$  &$0.54_{-0.21}^{+0.26}$  \\
 \hline 
AT2019dsg* & $43.59_{-0.05}^{+0.09}$  &$4.26_{-0.01}^{+0.01}$  &$1.79_{-0.02}^{+0.02}$  &$0.25_{-0.11}^{+0.04}$  &$5.75_{-0.10}^{+0.06}$  &$0.98_{-0.15}^{+0.02}$  &$0.40_{-0.03}^{+0.11}$  \\
 \hline 
AT2020ysg* & $44.61_{-0.04}^{+0.05}$  &$4.22_{-0.01}^{+0.01}$  &$2.10_{-0.01}^{+0.01}$  &$1.04_{-0.07}^{+0.09}$  &$7.80_{-0.00}^{+0.00}$  &$0.74_{-0.11}^{+0.16}$  &$0.41_{-0.21}^{+0.31}$  \\
 \hline 
AT2020yue* & $43.94_{-0.02}^{+0.02}$  &$3.96_{-0.01}^{+0.01}$  &$2.11_{-0.02}^{+0.02}$  &$0.40_{-0.32}^{+0.30}$  &$7.29_{-0.34}^{+0.30}$  &$0.78_{-0.14}^{+0.14}$  &$0.83_{-0.18}^{+0.13}$  \\
 \hline 
AT2021uqv* & $43.43_{-0.01}^{+0.01}$  &$4.14_{-0.00}^{+0.00}$  &$1.74_{-0.02}^{+0.02}$  &$0.47_{-0.12}^{+0.12}$  &$6.36_{-0.28}^{+0.35}$  &$0.74_{-0.23}^{+0.17}$  &$0.84_{-0.16}^{+0.11}$  \\
 \hline 
AT2022dsb* & $43.08_{-0.05}^{+0.05}$  &$4.92_{-0.21}^{+0.45}$  &$1.41_{-0.03}^{+0.03}$  &$-2.85_{-0.18}^{+0.18}$  &$7.39_{-0.30}^{+0.21}$  &$0.83_{-0.22}^{+0.13}$  &$0.29_{-0.07}^{+0.13}$  \\
 \hline 
AT2022hvp* & $44.97_{-0.02}^{+0.02}$  &$4.66_{-0.04}^{+0.04}$  &$1.45_{-0.01}^{+0.01}$  &$0.69_{-0.14}^{+0.13}$  &$6.84_{-0.40}^{+0.37}$  &$0.69_{-0.24}^{+0.20}$  &$0.82_{-0.22}^{+0.13}$  \\
 \hline 
AT2020mot* & $43.57_{-0.01}^{+0.01}$  &$4.22_{-0.01}^{+0.01}$  &$1.88_{-0.01}^{+0.01}$  &$0.54_{-0.19}^{+0.26}$  &$6.65_{-0.44}^{+0.45}$  &$0.56_{-0.30}^{+0.31}$  &$0.50_{-0.23}^{+0.24}$  \\
 \hline 
AT2018lni* & $44.03_{-0.10}^{+0.11}$  &$4.47_{-0.15}^{+0.33}$  &$1.85_{-0.03}^{+0.03}$  &$-0.90_{-1.29}^{+0.44}$  &$7.29_{-0.48}^{+0.29}$  &$0.46_{-0.26}^{+0.39}$  &$0.83_{-0.23}^{+0.13}$  \\
 \hline 
AT2021gje* & $44.27_{-0.02}^{+0.02}$  &$4.11_{-0.01}^{+0.01}$  &$1.95_{-0.02}^{+0.02}$  &$0.14_{-0.18}^{+0.15}$  &$7.68_{-0.02}^{+0.03}$  &$0.59_{-0.00}^{+0.00}$  &$0.04_{-0.02}^{+0.05}$  \\
 \hline

    \end{tabular}
    \caption{Same as \cref{tab:one power law}, but using the magnetized disk model with early-time exponential decay (Eq. \ref{eq: early model exp}). TDEs marked with an asterisk are those for which the model does not provide a good fit  (see \S\ref{subsec: mag fitting}).}
    \label{tab:mag exp}
\end{table}

\begin{table}[]
    \centering
    \begin{tabular}{|c|c|c|c|c|c|c|c|c|}
  
\hline 
TDE Name & $\log(\nu L_{\rm peak})$ & $\log(T_{\rm early})$ & $\log(t_0)$ & $p_{\rm decay}$ & $\log(\alpha)$ & $\log(M)$ & $\log(m_\star)$ & $\cos(i)$ \\ 
 \hline 
 & $\log(\rm{erg}/\rm{s})$ & $\log(\rm{K})$ & $\log(s)$ & $1$ & $1$ & $\log(M_\odot)$ & $\log(M_\odot)$ & $1$ \\ 
 \hline 
AT2019ahk & $43.85_{-0.01}^{+0.01}$  &$4.20_{-0.01}^{+0.01}$  &$1.97_{-0.07}^{+0.07}$  &$2.42_{-0.21}^{+0.25}$  &$-1.48_{-0.31}^{+0.28}$  &$6.81_{-0.49}^{+0.33}$  &$-0.17_{-0.20}^{+0.35}$  &$0.71_{-0.34}^{+0.20}$  \\
 \hline 
AT2020opy & $44.03_{-0.01}^{+0.02}$  &$4.20_{-0.01}^{+0.01}$  &$2.06_{-0.09}^{+0.09}$  &$1.91_{-0.26}^{+0.32}$  &$0.05_{-2.08}^{+0.34}$  &$6.31_{-1.29}^{+0.63}$  &$-0.13_{-0.61}^{+0.43}$  &$0.71_{-0.36}^{+0.21}$  \\
 \hline 
AT2020vwl & $43.68_{-0.06}^{+0.07}$  &$4.27_{-0.01}^{+0.01}$  &$1.48_{-0.08}^{+0.08}$  &$1.78_{-0.08}^{+0.10}$  &$-3.51_{-0.28}^{+0.34}$  &$6.47_{-0.50}^{+0.31}$  &$-0.71_{-0.21}^{+0.35}$  &$0.67_{-0.32}^{+0.24}$  \\
 \hline 
ASASSN-14ae & $43.68_{-0.02}^{+0.02}$  &$4.25_{-0.01}^{+0.01}$  &$1.96_{-0.07}^{+0.04}$  &$4.49_{-0.47}^{+0.35}$  &$-1.24_{-0.41}^{+0.42}$  &$6.67_{-0.63}^{+0.52}$  &$-0.08_{-0.29}^{+0.45}$  &$0.16_{-0.08}^{+0.14}$  \\
 \hline 
AT2018hyz & $43.95_{-0.01}^{+0.01}$  &$4.20_{-0.01}^{+0.01}$  &$1.80_{-0.06}^{+0.07}$  &$2.08_{-0.18}^{+0.25}$  &$-0.34_{-0.88}^{+0.62}$  &$6.56_{-0.67}^{+0.52}$  &$-0.14_{-0.29}^{+0.43}$  &$0.44_{-0.24}^{+0.29}$  \\
 \hline 
AT2023mhs & $44.10_{-0.05}^{+0.04}$  &$4.12_{-0.01}^{+0.01}$  &$1.75_{-0.07}^{+0.08}$  &$3.14_{-0.23}^{+0.32}$  &$-2.52_{-1.14}^{+1.07}$  &$6.14_{-0.94}^{+0.60}$  &$-0.44_{-0.42}^{+0.40}$  &$0.65_{-0.29}^{+0.24}$  \\
 \hline

    \end{tabular}
    \caption{Same as \cref{tab:one power law}, but using the magnetized disk model with early-time power-law decay (Eq. \ref{eq: early model PL}). TDEs marked with an asterisk are those for which the model does not provide a good fit  (see \S\ref{subsec: mag fitting}).}
    \label{tab:mag PL}
\end{table}

\end{document}